\documentclass[a4paper,11pt]{article}
\pdfoutput=1 

\usepackage{jheppub} 

\graphicspath{{Figures/}}
\usepackage{amsmath}
\newcommand{\beq}{\begin{equation}}
\newcommand{\eeq}{\end{equation}}
\newcommand{\tr}{\mathrm{Tr}}
\usepackage[dvipsnames]{xcolor}

\title{What can we learn about islands and state paradox from quantum information theory?}
\author[a,\dag]{Xuanhua Wang,} \note[\dag]{Corresponding authors.}
\author[b]{Kun Zhang,} 
\author[a,b,\dag]{Jin Wang}
\affiliation[a]{Department of Physics and Astronomy, Stony Brook University, Stony Brook, NY 11794, USA,}
\affiliation[b]{Department of Chemistry, Stony Brook University, Stony Brook, NY 11794, USA}
\emailAdd{xuanhua.wang@stonybrook.edu}
\emailAdd{kun.h.zhang@stonybrook.edu}
\emailAdd{jin.wang.1@stonybrook.edu}

\abstract{Recent discovery of the fine-grained entropy formula in gravity succeeded in reconstructing the Page curves that are compatible with unitary evolution. The formula of generalized entropy derived from the gravitational path integration, nevertheless, does not provide concrete insight on how the information comes out from the black hole given that the state of the radiation seems to follow what was given by Hawking. In this paper, we start from a qubit model and provide a quantum informational interpretation of entanglement islands and draw the parallel between the black hole information paradox and the problems of measurements in quantum mechanics. We show that the Page curve can still be realized even if information is lost and the information paradox can be attributed to the measurement problem. We argue that such interpretation is necessary for a quantum informational model if smooth horizons and bulk reconstruction are assumed, and show how it explains the Page curves of solvable models of 2D gravity. Though speculative, the similarities between the black hole information problem and the measurement problem may suggest some link in the origins of the two fundamental issues of distant fields.}

\begin{document} 
\maketitle
\flushbottom

\section{Introduction}
The information loss of black holes has been a long-standing problem that indicates incompatibility of several important fields of physics---quantum mechanics, thermodynamics and the theory of general relativity. It is remarked by Giddings that this paradox may be of equal importance to the ultraviolet crisis in classical physics which was only resolved by the advent of quantum mechanics \cite{Giddings:2006sj}. Nevertheless, people have shown widely different attitudes towards this problem mainly due to the lack of experimental verification and also the large number of proposals trying to resolve this issue. In this paper, we do not attempt to resolve this issue by adding exotic constructions, instead we incorporate recent discoveries of entanglement islands and give a quantum informational interpretation of what (effectively) has to occur inside the black hole based only on all we already know about quantum mechanics. 

Though tremendous progress has been made in the past two years to provide a quantum description of the black hole fine-grained entropy in the process of evaporation, the fine-grained entropy formula is still viewed as a black box which gives the expected answer to the important question but without telling us exactly how such result is obtained. It has been argued that including the quantum Ryu–Takayanagi (RT) prescription does not completely resolve the information problem in the bulk and it still requires substantial failure of semi-classical gravity such as firewalls \cite{Banks:2010zn,Bousso:2019ykv}. By now, many Page curves that conserve fine-grained entropy were explicitly calculated using the generalized entropy formula. The consistency of these results suggested that one should take the formula seriously even though there are still ambiguities in its derivation regarding the measure and the meaning of the gravitational path integral (GPI) \cite{Gibbons:1976ue,Hawking:1978jz}. As we mentioned, despite its success in providing reasonable answers, the formula did not hint on {\it how} the information comes out from the black hole or what the density matrix of the radiation is. The fine-grained entropy for the first half of the evaporation is dominant by the Hawking saddle (trivial RT surface). It suggests that the density matrix is consistent with that given by Hawking. At later times, calculations show that the entropy of black hole has an $\mathcal{O}(1)$ correction due to the dominance of the replica wormhole saddle. The decrease of entropy indicates that the information inside the horizon at late times becomes somehow encoded in the radiation outside. But no such mechanism is known and the state of the radiation is ambiguous. There are claims such that any further progress must involve a complete theory of quantum gravity. However, we do not subscribe to this viewpoint. For any specific quantum system, no ambiguity of state is allowed. According to Bekenstein-Hawking unitarity \cite{Marolf:2020rpm}, if the black hole evolves unitarily, then it obeys the laws of quantum mechanics and can be modeled as a quantum system. If any nonunitary behaviors show up in the evaporation of black holes, it is reasonable to assume that it can be approximated by some nonunitary operators in quantum mechanics. In short, we should be able to find a quantum informational model that effectively describes the behaviors of the black hole from the viewpoint of the outside. If not, then quantum mechanics may need to be heavily modified even in the asymptotically weak gravity limit which is not what we have observed in the lab. Such endeavor of building quantum mechanical models has been undertaken by numerous studies. For a non-exhaustive list, see Refs.~\cite{Horowitz:2003he,Gottesman:2003up,Harlow:2014yoa,Lloyd06,Lloyd:2013bza,Lloyd:2004wn,Bousso:2013ifa,Osuga:2016htn,Chatwin-Davies:2015hna,Borsten:2012fx,Akil:2018iny}. On the other hand, the new results from the last two years have boosted our understandings about the black hole but without yet providing a clear quantum informational interpretation. This paper aims to fill the gap and attempt a description of the island using quantum informational language.

In addition, there has been exuberant discussions on the smoothness of the horizon after the AMPS firewall argument \cite{Almheiri:2013hfa,Almheiri:2012rt,Bousso:2012as,Abedi:2016hgu,Polchinski:2016hrw}.\footnote{It is worth noticing that classical solutions of general relativity which resemble firewalls were also constructed explicitly \cite{Kaplan:2018dqx,McManus:2020lgm}. However, those firewalls are classical objects representing the Planckian shell of energy density near the horizon and are not solutions of black holes.} The logic for why firewall looks so unavoidable can be summarized as the following. {\it If} we assume that the whole evaporation process is unitary, then information must either come out or remain in certain forms of remnants or baby universes. We constrain our discussion to the first scenario which is suggested by the quantum RT results. For a review of remnant models, see Ref.~\cite{Chen:2014jwq}. If information is able to come out from the black hole, the radiation cannot be in the Hawking form, or equivalently, the vacuum near the horizon cannot be in the Unruh state which allows no outgoing information. Either unitarity or Hawking's semi-classical treatment breaks down \cite{Braunstein07}. If the vacuum near the horizon is not Unruh state, freely falling observers will not see vacuum but a wave of radiation at the horizon. Therefore, we will end up with the firewall scenario once requiring the unitary evolution of black holes and the validity of the effective field theories (ETFs) in the weak gravity limit. Assuming that a certain quantum gravity theory can magically resolve all the conflicts, it will still result in the ambiguity of the density matrix of Hawking radiations, which is termed as state paradox \cite{Bousso:2020kmy,Bousso:2019ykv}. Nevertheless, there are two loopholes in this line of argument. One is in the initial assumption. Quantum mechanics not only has the unitary evolution as the basic axiom but also has to include measurements which follow the Born rule. The second is that the outgoing radiation may not be in the same state when propagating from the horizon to the future infinity. For example, entangled Hawking pairs created near the horizon can gradually disentangle as they propagate away from the black hole. Such process is achievable by either nonunitary operations inside the black hole which can be described through Reeh-Schlieder theorem or nonlocal interactions of gravity \cite{Giddings:2006sj,Giddings:2006vu,Osuga:2016htn}.\footnote{As is remarked in \cite{Giddings:2006sj}, studies of high-energy scattering in string theory has not shown evidence of nonlocality on scales that would correspond to production of the requisite long strings despite that such nonlocality may resolve the information paradox of black holes.}  The firewall argument arises as a consequence of the disentanglement of Hawking pairs near the horizon. This argument indicates that something singular has to occur near the horizon. The gravitational fine-grained entropy formula offers an answer other than the energy curtain---the island. One should note that both island and firewall develop near the horizon after Page time, and both are singular objects, either in the energetic or the entropic sense, coined to avoid the information paradox. 
Therefore, it is desirable to have either firewall or island in one description of the evolution of black holes but not both. 

The fundamental inputs in the quantum mechanics not only include the unitarity of evolution but also the axioms of measurements. It came as a surprise that not all measurements in quantum mechanics can be understood as some local and unitary processes such as decoherence. The related problems are summarized as the measurement problem or Wigner's friend paradox \cite{wigner1995remarks,deutsch1985quantum,bong2020strong,schlosshauer2005decoherence,Bousso:2013uka}. To be able to describe all measurements on any quantum systems, it appears that either unitarity or locality in the process needs to be abandoned similar to the black hole information problem. Possible measurement-like processes inside black holes have been given serious considerations in Refs.~\cite{Bousso:2013uka,Lloyd:2013bza,Pasterski:2020xvn}. It draws our attention that the long-standing problem of measurement appears to be closely related to the black hole information problem as they are facing similar conundrums.\footnote{It has been believed that nonlocality is an intrinsic feature of strong gravitational dynamics \cite{Giddings:2006sj,Giddings:2006vu,Giddings:2004ud,Giddings:2005id}. This is doubtful if we consider the measurements on entangled quantum systems as certain forms of operators. Local phases can disappear in various decoherence programs, but global phases in entangled systems cannot be washed out from local interactions  \cite{schlosshauer2005decoherence}.} To have an overall unitary evolution of the black hole from the perspective of an outside observer, we seem to be forced to include either some nonunitarity inside the black hole so that it cannot be directly observed from the outside, or some nonlocal interactions of gravity. Similar issues also appear in the measurement problem. We will explain this point in more details in Sections~\ref{Subsec:implication_q_tele} and \ref{Subsec:crux}. 

In this paper, we provide one possibility to map the black hole information problem to the quantum measurement problem. We attempt at an effective description of the islands and Page curves within the standard quantum mechanics and information theory with least extra assumptions such as ER=EPR or traversable wormholes.\footnote{In this study, we argue that the effect of replica wormholes is indistinguishable from certain measurements at the horizon, therefore, ``ER"=``QT" where ``QT" stands for quantum teleportation.} It appears to the authors that the description may be the only option if one wants to describe black hole quantum mechanically while at the same time assumes the bulk reconstruction of the entanglement wedge and smooth horizons. We show that in this interpretation, the island is identified as a quantum measurement or a projection operator. The ``measurement" should be understood as effective description of certain complicated processes such that we may not be able to tell if it is fundamentally one basic operator or an emergence of a sum of simple operators. By this identification, we can restore the Page curves from a pure quantum mechanical argument. We discuss two possibilities which are indistinguishable from the outside that both give the same Page curve. The first possibility is a unitary evolution which conserves information and the second one loses information. For all practical purposes, the two scenarios look almost identical from the outside and therefore we cannot tell which one is the correct. The information loss in the second possibility is due to that the final states, though pure, are not in one-to-one correspondence or bijection with initial conditions. Thus we cannot reverse time to retrieve the initial data from a final state. This may result in an interesting scenario. The first possibility is what is expected from unitarity, while in the second case the final-state Hawking radiation is an ensemble of different pure states if we assume that the measurement is nonunitary. In this proposal, the state of radiation is well-defined and the idea of gravity/ensemble duality has natural realization. However, as to how such measurement (projection) occurs, we suggest a connection to the quantum-to-classical transition but we do not pretend to have an definite answer for it \cite{kofler2007classical,penrose1996gravity,bassi2013models}.

The paper is arranged as follows. In Sec.~\ref{Sec:gen} we provide the background on state transfer protocols and give a general explanation of the proposal and the rationale behind. The proposal is summarized in a dictionary. In Sec.~\ref{Sec:related}, some existing quantum informational models and ideas related to our proposal are reviewed and discussed. In addition, we comment on the implications of our proposal on the quantum no-cloning theorem in black holes. In Sec.~\ref{Sec:2Dmodel} we review the issues related to the purification rates in 2D dilaton gravities and provide a prescription for the problem within our proposal. In Sec.~\ref{Sec:discussion}, we summarize our proposal and give a general account of the ramifications, the remaining issues and the alternative approaches.


\section{Wormholes or emergent measurements?}\label{Sec:gen}
The inside of a black hole is tumultuous \cite{Belinsky:1970ew}, and therefore it is not unreasonable to assume that some processes have an effective description in terms of measurement, regardless of its intrinsicality. In Subsection \ref{Subsec:motiv}, we gives the motivation and why a quantum informational model is necessary for our understanding. Subsection \ref{Subsec:swapping} describes the state transfer protocols, realized by quantum entanglement and measurements, namely the teleportation and entanglement swapping. Subsection \ref{Subsec:implication_q_tele} discusses the implications of those protocols, especially the issues about quantum measurements. Subsection \ref{Subsec:crux} discusses the relation between the protocols and the black hole information. The correspondence of island and measurement is proposed. In the last subsection, draw the link from the related concepts in gravity theory to those in quantum information and summarize them in an {\it ad hoc} dictionary.

\subsection{Why purification is necessary}\label{Subsec:motiv}
It is argued that early Hawking radiation gets purified due to the nonperturbative corrections to the density operator 
\begin{align}
    \rho_{\rm exact}=\rho_{\rm thermal}+{\rm perturbative}+\mathcal{O}(e^{-\# S_{\rm BH}})\,,
\end{align}
where the last term is the nonperturbative correction due to, for example, the appearance of the replica wormholes. The perturbative correction does not fix the information problem when considering the massive black holes \cite{Harlow:2014yka}. The nonperturbative correction which looks like semi-random noise that is exponentially suppressed, gives $\mathcal{O}(1)$ corrections to the entanglement entropy of the black hole after a sufficient amount of radiations have been emitted \cite{Papadodimas:2013kwa,Papadodimas:2012aq,Stanford:2020wkf}. For a massive evaporating black hole, calculations using the generalized entropy formula suggest that after Page time, the outgoing Hawking particles have to be highly entangled with the early radiation so that the decrease rate of the radiation entropy is roughly the same order as the increase in the initial period. This requires $\mathcal{O}(1)$ corrections to the density matrix of the micro-process which results in a ``firewall" at the horizon. 

The quantum RT calculation gives Page curves that decay to zero in the final stage, however, the state of the radiation becomes ambiguous when adopting the holographic arguments \cite{Bousso:2019ykv,Bousso:2020kmy}. If a futuristic experiment were to be conducted to collect all Hawking radiations from a lab-made black hole, a pure state has to be found in order to have the Page curve. The interpretation of island seems to obscure the radiation density matrix as the information within the island is understood as encoded in the radiation, but the experimental result only gives a yes or no answer and cannot be ambiguous. The density matrix can in principle be reconstructed through quantum tomography if sufficient identical copies of the states are given \cite{NC10}. Though the physical meaning of the islands which are the mouths of the Euclidean replica wormholes is unclear, one should note that the replica wormholes appear in the middle of the calculation of the entanglement entropy. The result of the calculation is clear---standing away from the black hole, the early radiation, which can be described by standard quantum mechanics, is seen to be purified by the late radiation and the total radiation entropy decreases. Therefore, we will see photons entangled with early Hawking photons come into our delicate experimental apparatus and this is an order-one change to the results from effective theories. We do not intend to modify the applicability of EFT, instead we use only quantum mechanics and show what {\it effectively has to occur} inside the black hole if the Page curves are approximately correct.

\subsection{Background on quantum teleportation and entanglement swapping protocols} \label{Subsec:swapping}
Quantum teleportation and entanglement swapping protocols  describe how to transmit quantum states with the help of quantum entanglements and measurements \cite{Bennett93}. Suppose that the one qubit (as pure state) has the generic form
\begin{equation}
    |\varphi\rangle = \alpha|0\rangle+\beta|1\rangle,
\end{equation}
where the amplitudes $\alpha$ and $\beta$ represent the encoded information. The task is to transmit $|\varphi\rangle$ state from Alice to Bob, where the state $|\varphi\rangle$ is also unknown to Alice. Instead of physically sending the qubit to Bob, teleportation can send the information of state $|\varphi\rangle$ without transmission any qubits. 

The prerequisite for teleportation is the shared entangled state (Bell state) between Alice and Bob, denoted as
\begin{equation}
    |\Phi^+\rangle_\text{AB} = \frac{1}{\sqrt 2}(|0\rangle_\text{A}\otimes|0\rangle_\text{B}+|1\rangle_\text{A}\otimes|1\rangle_\text{B}).
\end{equation}
Then we can rewrite the initial state $|\varphi\rangle_\text{A}\otimes |\Phi^+\rangle_\text{AB}$ in the Bell basis $\left(1\!\!1_2\otimes \sigma^j_x\sigma^k_z\right)|\Phi^+\rangle$ (with $j,k=0,1$) of $\mathcal H_2\otimes \mathcal H_2$ as follows,
\begin{equation}
\label{eq:tele_eq}
    |\varphi\rangle_\text{A}\otimes |\Phi^+\rangle_\text{AB} = \frac 1 2 \sum_{j,k=0}^1 \left(1\!\!1_2\otimes \sigma^j_x\sigma^k_z\right)|\Phi^+\rangle_\text{AA}\otimes \sigma^j_x\sigma^k_z|\varphi\rangle_\text{B},
\end{equation}
where $\sigma_{x(z)}$ are the Pauli matrices. We also take the notations for the other Bell states: $|\Phi^-\rangle = (1\!\!1_2\otimes\sigma_z)|\Phi^+\rangle$, $|\Psi^+\rangle = (1\!\!1_2\otimes\sigma_x)|\Psi^+\rangle$, $|\Psi^-\rangle = (1\!\!1_2\otimes\sigma_x\sigma_z)|\Psi^+\rangle$. After the measurement, one can see that the information about the state $|\varphi\rangle_A$ is transferred to $|\varphi\rangle_B$. In fact, Eq. (\ref{eq:tele_eq}) does not represent the teleportation process, instead it is only the algebraic trick to rewrite the state $|\varphi\rangle$ from the location A to the location B. The RHS of Eq. (\ref{eq:tele_eq}) is in a superposition form. Measurement can eliminate the superpositions. Performing the projector $|\Phi^+\rangle_\text{AA}\langle \Phi^+|$ on the initial state, which gives
\begin{equation}
\label{eq:tele_eq2}
    |\Phi^+\rangle_\text{AA}\langle \Phi^+|\varphi\rangle_\text{A}\otimes |\Phi^+\rangle_\text{AB} = \frac 1 2 |\Phi^+\rangle_\text{AA}\otimes |\varphi\rangle_\text{B}.
\end{equation}
The normalization factor $1/2$ suggests that the measurement result $|\Phi^+\rangle_\text{AA}$ has the probability $1/4$. The projector $|\Phi^+\rangle_\text{AA}\langle\Phi^+|$ ``forces'' Bob's state to be $|\varphi\rangle$. Other measurement results (projection onto other Bell states) will transform Bob's state to be $|\varphi\rangle$ up to some Pauli operations. Note that the no-cloning theorem is not violated \cite{Wootters82}, since the measurement results does not reveal any original information of state $|\varphi\rangle$.

The project measurement is the key in teleportation process, but it is not necessary to be the Bell project measurement. One can check that a more general teleportation operation than Eq. (\ref{eq:tele_eq}) that includes a unitary operation can be written as
\begin{equation}
\label{eq:U_AA}
    (U_\text{AA}\otimes 1\!\!1_\text{B}) |\varphi\rangle_\text{A}\otimes |\Phi^+\rangle_\text{AB} = \frac 1 2 \sum_\text{j,k=0}^1 |\psi(jk)\rangle \otimes \sigma^j_x\sigma^k_z|\varphi\rangle_\text{B},
\end{equation}
with $|\psi(jk)\rangle =  U_\text{AA} \left(1\!\!1_2\otimes \sigma^j_x\sigma^k_z\right)|\Phi^+\rangle_\text{AA}$. States $|\psi(jk)\rangle$ also form an orthonormal basis, which can be the product state basis $|jk\rangle$, if the corresponding unitary transformation $U_\text{AA}$ is applied. 


Teleportation is not limited to the pure state transfer. In the original proposal of teleportation \cite{Bennett93}, the authors argue that Alice can also teleport a mixed state to Bob. Suppose that the qubit (given to Alice) prepared to be teleported is completely mixed, i.e., $\rho_\text{A} = 1\!\!1_2/2$, which has the purification of the maximal entangled state, such as $|\Phi^+\rangle_\text{CA}$. Then the teleportation equation (\ref{eq:tele_eq}) can be generalized to
\begin{equation}
\label{eq:es_eq}
   |\Phi^+\rangle_\text{CA} \otimes |\Phi^+\rangle_\text{AB} = \frac 1 2 \sum_{j,k=0}^1 \left(1\!\!1_2\otimes \sigma^j_x\sigma^k_z\right)|\Phi^+\rangle_\text{AA}\otimes \left(1\!\!1_2\otimes \sigma^j_x\sigma^k_z\right)|\Phi^+\rangle_\text{CB}.
\end{equation}
Similarly, to pick up one specific state on the RHS, we can perform the projector $|\Phi^+\rangle_\text{AA}\langle \Phi^+|$, which gives 
\begin{equation}
    |\Phi^+\rangle_\text{AA}\langle \Phi^+|\Phi^+\rangle_\text{CA} \otimes |\Phi^+\rangle_\text{AB} = \frac 1 2 |\Phi^+\rangle_\text{AA}\otimes |\Phi^+\rangle_\text{CB}.
\end{equation}
In other words, the teleported mixed state preserves its correlation to the other state. Then the original correlation between CA has swapped to CB. Any unitary evolution $U_\text{AA}$ can be applied before the teleportation, then the project two-qubit basis has to been changed accordingly. If the projection basis does not match the unitary evolution $U_\text{AA}$, the fidelity of the teleported state is less than 1 and we cannot fully recover the initial state $|\varphi\rangle$.

\subsection{Implications of teleportation and entanglement swapping} \label{Subsec:implication_q_tele}

If Bob does not know Alice's measurement result, such as $|\Phi^+\rangle_\text{AA}$, Bob can only describe his state as an ensemble, which looks like a mixed state for him. However, in the viewpoint of Alice, her measurement purifies Bob's state. We should distinguish the ignorance of the mixed state to be classical or quantum. Mixed state due to classical ignorance, like Bob's mixed state description, does not contribute to the entanglement entropy. Since entanglement entropy only counts the local ignorance due to the entanglement, which is beyond the classical description, such as the local hidden variable theory.

Although the teleportation and entanglement swapping protocols involve the nonunitary measurements, there is no information loss. Suppose that the information is encoded locally in the qubit $|\varphi\rangle$ or nonlocality in the entanglement. The teleportation and entanglement swapping can always restore the original unknown information if the measurement results are accessible. Therefore, teleportation and entanglement swapping are referred as the information transfer protocols, which do not have destructive effects on quantum information. However, this is on the assumption that the initial state is not randomly chosen and can only be in one of the maximally-entangled forms. Otherwise, the fidelity between the initial and final states is less than 1 and part of the information will be lost.

Teleportation and entanglement swapping take advantage of the state reduction (wave function collapse) in measurement processes, which is beyond the decoherence picture. Eqs. (\ref{eq:tele_eq}) and (\ref{eq:es_eq}) clearly show that one can always rewrite state $|\varphi\rangle_\text{A}$ at location A as superpositions of state $|\varphi\rangle_\text{B}$ at location B with the help of quantum entanglement. Quantum measurement picks up one state from superpositions, therefore teleportation can happen. If one views the measurement process as a unitary evolution between the measured system and the measurement apparatus with state $|m\rangle_\text{M}$ (like decoherence), then the teleportation process can be described as $\left(U_\text{MAA} \otimes 1\!\!1_\text{B} \right) |m\rangle_\text{M}\otimes |\varphi\rangle_\text{A}\otimes |\Phi^+\rangle_\text{AB}$ with the measurement evolution $U_\text{MAA}$. Therefore, Bob's state is given by
\begin{equation}
    \tr_\text{MAA} \left[\left(U_\text{MAA} \otimes 1\!\!1_\text{B} \right) |m\rangle_\text{M}\langle m|\otimes |\varphi\rangle_\text{A}\langle \varphi|\otimes |\Phi^+\rangle_\text{AB}\langle \Phi^+|\left(U^\dag_\text{MAA} \otimes 1\!\!1_\text{B} \right)\right] = \frac 1 2 1\!\!1_\text{B},
\end{equation}
which is not transferred to $|\varphi\rangle_B$ in the view of Alice or Bob. Besides, Bob's state is not purified and still entangled with Alice's state (plus the measurement apparatus). Any unitary evolution between Alice's state and the measurement apparatus does not change Bob's state. 

The measurements in teleportation and entanglement swapping has a nonlocal effects (also true for measurements on any spatial separated entangled states), which can be explicitly seen through quantum tomography. The nonlocal phenomenon has a long debate in history and is still ongoing.\footnote{As phrased by Schr\"odinger \cite{schrodinger35}: ``It is rather discomforting that the theory should allow a system to be steered or piloted into one or the other type of state at the experimenter's mercy in spite of his having no access to it.''} Such phenomenal has been coined as quantum steering in quantum information science \cite{Uola20}. Note that teleportation or entanglement swapping does not violate the no-signaling principle, also known as the no-communication theorem. Signaling is forbidden by the total random measurement results. Both teleportation and entanglement swapping have been experimentally verified \cite{Bouwmeester97,Pan98}, even between the space-like region \cite{Herbst15}. We emphasis that if the nonunitary measurement is viewed as a part of unitary operation (like the evolution of open quantum systems), the measurements in teleportation and entanglement swapping have to be a part of global unitary evolution acting on the space-like region. Such description violates the causality, which is not less bizarre than the wavefunction collapse. 

Another striking feature about entanglement swapping is the ambiguous time when the entanglement is swapped \cite{ma2012experimental,Ma:2014dta}. More precisely, recent experiments have shown that the entanglement can be swapped before even the measurement occurs. In other words, we can test the entanglement properties of the system before the measurement operations. Such 'steering into the past' behavior puts challenges on the understanding of the wavefunction if the strict causality is posed. Though this may have further implications on the entanglement structure, the discussion is beyond the scope of this paper. 

\subsection{Island--measurement correspondence} \label{Subsec:crux}
The crux of the proposal is the following. To have a quantum informational description of the entanglement wedge inside the black hole, the island is to be identified as where information is acausally transferred to the outside. Consider a black hole which initially has a low entropy and a trivial quantum RT surface. The infalling Hawking state is maximally entangled with the outgoing state assuming microcanonical description and this add to the entropy of the black hole. When the black hole is sufficiently mixed, it operates as a measurement apparatus or projection operator onto the Unruh state with an additional black hole state. The assumption that information will be teleported out is equivalent to a predictable ``measurement" result assuming one has a complete knowledge about the apparatus states and the state to be measured. For a black hole with fine-grained entropy equal to Bekenstein-Hawking entropy, the infalling negative energy state and a black hole state need to be projected onto the vacuum state near the horizon, similar to an annihilation process seen from the outside. The black hole state will be transferred to the radiation through an entanglement swapping protocol, with the classical information channel replaced by a complete theory of measurement which allows one to calculate the measurement results. Therefore, the fine-grained entropy of the black hole (radiation) decreases. This process occurs right at when the island forms near the horizon. On the other hand, one should note that the island is derived from the replica calculation of entanglement entropy. It does not directly imply the transfer of information, more strictly, it is interpreted as where the decrease of the entanglement entropy occurs. This is equivalent to the purification of Hawking radiation. If we do not require the information to come out from the black hole, we only need to impose a Bell-state measurement on the island without the additional information transmission, or the predetermined experiment results. The radiation is purified in the fine-grained sense. In this case, the black hole (radiation) has the same Page curve as that in the previous case. However, the outgoing radiation is an ensemble of pure states and we cannot retrieve any information from it. The merit of this idea is that it is entirely within the frame of quantum mechanics, and no postselection or new postulates are needed to obtain the correct Page curve. The assumption we do require is that one can treat the black hole as a semi-classical and macroscopic objects where measurement-like processes in quantum mechanics can also happen there on the horizon depending on the state of the black hole.

\begin{figure}[ht]
    \centering
    \includegraphics[width=0.38\textwidth]{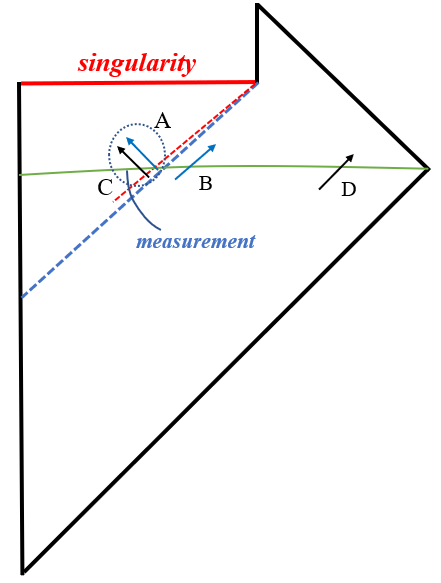}
    \caption{The Hawking pair AB marked in blue created near the horizon (A and B are conventionally drawn parallel to the horizon as represented by the Unruh vacuum. Here, the picture is drawn to represent the particle-hole picture). The ingoing qubit A reaches the island represented by the slanted red line and gets measured. The Cauchy surface chosen at the instant of measurement is the green horizontal line. The early radiation D in black is maximally entangled with the black hole state C.}
    \label{a}
\end{figure}

The scrambling time has a simple explanation in terms of projection operations. The scrambling time was originally argued in a setup where a black hole directly spits out its information (qubits) as the Hawking radiation after Page time and information thrown into the black hole can be extracted out from radiation after a scrambling time \cite{Hayden:2007cs,Sekino:2008he}. From the dictionary of AdS/CFT, the bulk scattering processes in the causal wedge $\mathcal{C}_A$ of $A$ are reconstructable in terms of certain nonlocal CFT operators in $A$ through Hamilton-Kabat-Lifschytz-Lowe (HKLL) procedure \cite{Hamilton:2006az}
\begin{align}
    \langle \phi(x_1)\phi(x_2)\rangle_g=\int dX_1 dX_2\ K(x_1;X_1) K(x_2;X_2) \left\langle 0|\mathcal{O}(X_1)\mathcal{O}(X_2)|0\right\rangle_{CFT}\,,
\end{align}
where $K(x_i;X_i)$ is the smearing operator which is dependent on the specifics of the sapcetime. This understanding of causal wedge reconstruction was later proposed to extend into the larger entanglement wedge $\mathcal{E}_A$ \cite{Czech:2012bh}. In our case, it suggests intuitively that when a qubit enters into the entanglement wedge of radiation, the state of the qubit can be reconstructed from the radiation outside the horizon. In the quantum informational language, when the qubit enters the island, its state will be teleported out through the aforementioned protocols. This is similar to the Horowitz-Maldacena (HM) proposal \cite{Horowitz:2003he}. The difference is that in HM model, a pre-determined projection into one particular state is introduced for all possible incoming states at the singularity, and the model does not consider the emergence of entanglement wedge after Page transition. Therefore, from the HM proposal a trivial Page curve is obtained. We will discuss this proposal in more detail in Sec.~\ref{Subsec:HM}. Before the island forms, no such measurements occur and the entropy increases monotonically as Hawking predicted. As we discussed, the measurement can be described by either a nonunitary process which still gives the same Page curve but has information loss, or a nonlocal unitary process which preserves information. Both interpretations of measurements exist in literature but it is disputable which one is more desirable.\footnote{One possible origin of the nonlocality in field theory comes from the holographic calculation. Commutators of spacelike separated fields, e.g. according to HKLL reconstruction, do not vanish, which violate the microcausality \cite{Hamilton:2006az}} This nonlocality in a measurement does not necessarily have to do with quantum gravity but it is directly related to the information loss problem in black holes. Such nonunitary or nonlocal effects are present in ordinary measurements of entangled and spacelike-separated quantum systems. One of the reasons why this perspective attracts us is that it links two different but essentially similar problems in black holes and in fundamental quantum mechanics. Both effects are suspected to be nonunitary in the respective fields which leads to information loss, and both involve the nonlocality as a way around. In addition, it seems that within the framework of quantum mechanics, the effective description has to be what is given in the paper based on our current understanding of black holes. 

In below, we will give an example to show how the above arguments work. The diagrammatic illustration of this mechanism in an asymptotically flat black hole background is given in Fig.~\ref{a}. We consider a Hawking pair $|R\rangle_{\text{in}\otimes\text{out}}$ created near the horizon slightly after Page time. The black hole state $|m\rangle$ is in maximal entanglement with early radiation $|i\rangle$. In the microcanonical description and assume that each qudit in the black hole is maximally entangled with the early radiations at the time of this event. Then the initial state in this micro-process
\begin{gather}
     |\Psi\rangle_\text{total initial}= \frac{1}{\sqrt{N}}\sum_i^{N}|m_i\rangle\otimes|i\rangle\otimes|R\rangle_{\text{in}\otimes\text{out}}= \frac 1 N \sum_{ij} |m_i\rangle\otimes |j\rangle_\text{in}\otimes |j\rangle_\text{out}\otimes|i\rangle
\end{gather}
can be simplified (without much loss of generality) to the qubit description\footnote{The four Bell states form a basis for the Hilbert space $\mathcal{H}_2\otimes\mathcal{H}_2$, therefore, they cannot all correspond to the vacuum. Here, we pick the singlet state as the vacuum state and the other three states then correspond to the firewall states.}
\begin{gather}
    |\Psi\rangle_\text{initial}=  |\Psi^-\rangle_\text{AB}|\Psi^-\rangle_\text{CD} = \frac{1}{2}(|0\rangle_A|1\rangle_B-|1\rangle_A|1\rangle_B)(|0\rangle_C|1\rangle_D-|1\rangle_C|1\rangle_D)\,,\label{eq:2entpair}
\end{gather}
where A and B are the infalling and outgoing Hawking states, and C is the black hole state that is maximally entangled with the early radiation state D. When the qubit A reaches the island, looking from the outside it ``annihilates" with a black hole state C. If we require the result of this annihilation to be a unique vacuum state, then it can be simulated as a projection into the maximally entangled vacuum state $|\Psi^-\rangle_{\rm AC}$. Notice that Eq.~\eqref{eq:2entpair} can be rewritten as
\begin{align}
     |\Psi\rangle_\text{initial}&= \frac{1}{2}|\Phi^+\rangle_\text{AC}|\Phi^+\rangle_\text{BD}-\frac{1}{2}|\Phi^-\rangle_\text{AC}|\Phi^-\rangle_\text{BD}\nonumber\\
     &-\frac{1}{2}|\Psi^+\rangle_\text{AC}|\Psi^+\rangle_\text{BD}+\frac{1}{2}|\Psi^-\rangle_\text{AC}|\Psi^-\rangle_\text{BD}\,,
\end{align}
which is the same as Eq. (\ref{eq:es_eq}). Then only the last term is postselected and the state is renormalized. The infalling state corresponds bijectively to the radiation state and no information is lost. On the other hand, if we do not postselect the state, the operation on the AC state is an ordinary measurement. No matter what the result of the measurement is, the radiation state turns from a mixed state to an ensemble of pure states and the fine-grained entropy decreases. Mathematically, this is due to that the ensemble average is done on the entropy $S_{vN}$ instead of on the density matrix, i.e. $\langle S_{vN}(\rho)\rangle=0$ while $S_{vN}(\langle\rho\rangle)\ne 0$. 

For the case that the island is identified as a standard quantum measurements, if the measurement is described by local unitary operators, then the quantum teleportation and entanglement swapping protocols will not exist in the first place. It is true unless one adopts the many-world interpretation, then the Hawking radiation in our universe is one pure state and there is an ensemble of different realities. This point is too philosophical and we do not go into that. If the measurement is an effective nonlocal unitary operation, then information will be conserved as the whole evaporation process is unitary. If the measurement is described as fundamental nonunitary and that the measurement result follows the Born rule, information will be lost inside the black hole.\footnote{Note that by ``information" here we mean the ability to reverse the process and retrieve the initial data, such as the initial state of black hole, rather than the difference between the maximal entropy and the fine-grained entropy.}  The projection model is one example of a nonunitary operation but its ``measurement" result does not follow the Born rule and thus information can be transmitted to the outside. One can argue that the Born rule is a statistical result and previous measurement results stored in the black hole may influence the next ``measurement" and at least part of the information can be retrieved from the outside. The differences between those models do not appear in the Page curve and are experimentally indistinguishable, thus we consider them as different interpretations of measurements. 

According to the Hayden-Preskill argument, any infalling information ``thermalizes" with the black hole information in a scrambling time. The information can be deciphered if Hawking radiation contains data about the black hole state instead of being in thermal state. It has been shown in various models that the scrambling time is approximately what it takes for the state to reach the island. 
Consider a Hayden-Preskill process where a dictionary which is maximally entangled with the observer is thrown into the black hole \cite{Hayden:2007cs}. The dictionary will be measured once entering the island and the information encoded in the entanglement will be {\it swapped} into the outgoing qubit $|B\rangle$ and the early radiation $|D\rangle$. If we adopt the projection proposal which requires a unique measurement result, then the information can be retrieved after in a scrambling time, consistent with the quantum informational analyses given in \cite{Hayden:2007cs,Sekino:2008he}. For a pure state thrown into the black hole, the projection generates quantum teleportation instead of entanglement swapping. However, if we release this requirement of a unique final state, information about initial conditions will be lost if the measurement results are not accessible. The information can be recovered if only the measurement result inside the black hole can be accessed from the outside. To conserve the information, certain violation of the randomness of the measurement is necessary assuming the horizon is smooth. Since the black hole does not conserve global symmetry charge, the measurement results may also disappear after the evaporation \cite{Kallosh:1995hi,Banks:2010zn,Hsin:2020mfa}. Alternatively, one can refer to the baby universe model where the information is stored, but the result is the same---if we measure the Hawking radiation from the outside, some information will be lost from our universe. 

In the measurement model, the uncertainty in the results of the measurements renders the density matrix of the outside as an ensemble. However, one should not confuse the coarse-grained density matrix where the ``mixed state" is the classical average due to our ignorance with the fine-grained density matrix, where mixed states originate from entanglement with other systems. Classical ensembles do not change the vanishing fine-grained entropy of the radiation and the Page curve. Besides, one can always expand a state in the Bell-basis. But a measurement process occurs when: $1.$ quantum coherence in the density matrix vanishes and $2.$ the diagonal terms in the measurement basis become different realities and are reinterpreted as classical probabilities in an ensemble. Though there are disputes on the interpretations of measurements,\footnote{The decoherence program has helped in the understanding of environment-system interactions and can explain some types of measurements \cite{schlosshauer2005decoherence}, but such mechanism does not resolve measurement problems related to the selection of one particular state and the violation of weak causality.} we want to emphasize that the issues regarding the states of Hawking radiation and the unitarity of quantum measurements are closely related, and we believe that a clear resolution of either problem can shed light to the other.

\subsection{Dictionary from gravity theory to quantum information}
We find that several quantities in black hole physics and in quantum information theory can be related by the following correspondence. 

{\it Entanglement island:} Bell-state measurement or projection.\\
The region inside the island, or the entanglement wedge of the radiation, is usually interpreted as where the information is encoded in the radiation. This understanding is borrowed from the bulk reconstruction in the AdS/CFT duality that bulk operators in AdS can be reconstructed as the CFT operators in spacial region $A$ provided that they lie in its entanglement wedge \cite{Dong:2016eik}. However, the fine-grained entropy formula is not dependent on the AdS/CFT correspondence and has been applied far beyond AdS black holes. In the language of quantum information, it is the region where the state information is transferred to the outside [details provided in Sec.\ref{Subsec:swapping}]. One important remark we want to draw is that this process cannot be realized through local unitary processes. Naively, we expect that the nonunitarity is a result of tracing the uninterested degrees of freedom and the whole system evolves unitarily. However, this cannot be true considering the realization of entanglement swapping. It is unclear if measurements conserve information and therefore the black hole information paradox can also be attributed to the old problem of measurement in quantum mechanics.

{\it Replica wormhole:} an entanglement swapping or a teleportation protocol. \\
It is not clear whether the island derived from the generalized entropy formula reflects anything physical or it is due to the out of control of the GPI. However, the general picture is as follows. The island represents the mouth of a replica wormhole. Information entering the island goes through the replica wormhole and returns to the asymptotic region. One related argument is the ER=EPR provided by Maldacena and Susskind \cite{Maldacena:2013xja}. On the other hand, one interpretation of the island which seems more natural in the framework of standard quantum systems is that the information is teleported out via certain teleportation-like protocols, i.e. ``ER=QT". More complicated multi-particle teleportation protocols can be identified as wormholes of higher genuses.

{\it Scrambling time:} the time before the measurement is completed.\\
This naturally follows from the above interpretation of the island and replica wormholes.

{\it Page transition:} quantum-to-classical transition or emergence of measurements.\\
Page transition occurs when island appears. In other words, it occurs when measurements effectively emerge. Therefore, Page transition can correspond to the boundary between the quantum and the classical and is described by certain quantum-to-classical transitions. The transition point can be inferred through the GPI defined below. 

{\it Gravitational path integral:} path integral in the many-world interpretation of measurements.\\
This is induced from the above parallels. Multiple trajectories can only exist as classical ensembles but quantum mechanically they exist simultaneously as a superposition. Universes of different geometries or topologies in QPI are like classical trajectories in standard quantum mechanics,
\begin{align}
    Z=\int Dg D\phi e^{-S_E[g,\phi]}Z_{matter}\,,
\end{align}
where 
\begin{align}
    S_E[g]=-\frac{1}{16\pi G}\int\sqrt{g}(R+...)+{\rm boundary}
\end{align}
is the Euclidean action of gravity. Semi-classically, the gravity part is only considered near the classical saddles of the general relativity $Z_{gr}\propto \sum_i e^{-S_{gr,i}}$. The many-world interpretation is conceptually similar that different classical realities coexist as a superposition. The Page transition corresponds to when the dominant geometry jumps from the Hawking saddle to the wormhole saddle based on the entropy. In terms of measurement, this corresponds to when the state of radiation jumps from the Unruh state to certain unentangled state. The inner product of the two state is exponentially suppressed and the limit of a large number of particles $N\rightarrow\infty$ returns the correct classical limit. In the gravity part, the classical saddles can be understood as certain ensemble average of a much larger set of quantum mechanical solutions which behave like random noise around the classical solutions. In the measurements which purify the early radiation, the state of the radiation after the measurements is an ensemble of pure states which has a sharp peak at a classical distribution. 

{\it Gravity/ensemble duality:} unitary interpretation/spontaneous wavefunction collapse indistinguishablity.\\
The gravity/ensemble duality is a proposal that the GPI calculates an ensemble of quantum mechanical theories without gravity \cite{Penington:2019kki,Stanford:2020wkf,Bousso:2020kmy,Bousso:2019ykv,Pollack:2020gfa}. This idea was proposed to resolve the state paradox of Hawking radiation and the factorization issue of the partition function in AdS/CFT and is supported by the results from random unitary dynamics. On the quantum information side, there are many interpretations of measurements. Unitarity can be restored in a quantum measurement assuming the certain superposition of classical realities. Such interpretation is indistinguishable from the nonunitary interpretation of the spontaneous collapse into a classical ensemble when the complexity of the system is large enough. The semi-classical gravity calculation suggests that a transition of dominant contributions to classical observables occurs when the fine-grained entropy of the measuring system is close to its thermal limit. This is the statement that the measuring system has to be classical.

In general, it is not yet completely clear how far we can go with the above correspondences. At the end of the evaporation when the black hole cannot be treated semi-classically, the effective theory of measurement is also assumed to break down. If the proposed connections are accurate, we may be able to learn about the fundamentals of quantum mechanics from the black hole, and vice versa. To gain a better understanding of the details and the accuracy of the dictionary, more work needs to be done.

\section{Quantum informational prescriptions and related ideas}\label{Sec:related}
In this section, we review some of the previous arguments that are related to our proposal and discuss the differences between these ideas and proposals. In Sec.~\ref{Subsec:HM} we review the HM proposal. We discuss its relation to our proposal and its shortcomings. In Sec.~\ref{Subsec:OP} we review the Osuga-Page (OP) model and comment on its strength and limitations. In Sec.~\ref{Subsec:RS} we comment on the connection between our model and the Reeh-Schlieder theorem. In Sec.~\ref{Subsec:no-cloning}, we review the no-cloning theorem and explain the quantum informational prescription according to this model. 

\subsection{Final-state projection of black holes}\label{Subsec:HM}
There has been extensive discussions on the HM proposal, which claims that the unitarity of black hole evaporation can be realized by imposing a final state boundary condition at the black hole singularity \cite{Horowitz:2003he}. The essential idea is similar to the quantum teleportation, but instead of allowing uncertainties in the outcome of the measurement inside the black hole, the outcome (the final state) is required to be definite to avoid the information loss. The gist of the argument is the following. In the semi-classical picture, the radiation is created in the form of Unruh state $|R\rangle$ (R stands for the radiation), given by
\begin{equation}
    |R\rangle_{\text{in}\otimes\text{out}} = \frac 1 {\sqrt{N}} \sum_j  |j\rangle_\text{in}\otimes |j\rangle_\text{out}\,.
\end{equation}
Here $|R\rangle$ is a maximally entangled two-qudit state where each qudit has dimension $N$. Qudit generalizes qubit into a $N$ level of state. 

The pure state which forms the black hole is denoted by $|\psi\rangle_\text{M}$. The black hole plus the infalling radiation and the outgoing radiation together form the closed system, given by $|\psi\rangle_\text{M}\otimes |R\rangle_{\text{in}\otimes\text{out}}$. Horowitz and Maldacena argued that the information loss arises because the singularity is treated as another ``asymptotic'' region. Therefore, no unitary operation can save the information from being lost. The key assumption in the HM proposal is that the singularity is \textit{not} another asymptotic region, instead, it is forced to be a unique pure state with the form
\begin{equation}
    |BH\rangle_{\text{M}\otimes \text{in}} = \frac 1 {\sqrt N} \sum_j |j\rangle_\text{M}\otimes |j\rangle_\text{in}\,,
\end{equation}
which is a maximally entangled two-qudit state. Both the Hilbert spaces $\mathcal H_\text{M}$ and $\mathcal H_\text{in}$ are considered as inside the black hole.

Analogous to Eq. (\ref{eq:tele_eq2}), this result is equivalent to a teleportation
\begin{equation}
    |BH\rangle_{\text{M}\otimes \text{in}}\langle BH|\psi\rangle_\text{M}\otimes |R\rangle_{\text{in}\otimes\text{out}} = \frac 1 N |BH\rangle_{\text{M}\otimes \text{in}} \otimes |\psi\rangle_\text{out}\,.
\end{equation}
The information about the initial state of the black hole $|\psi\rangle_\text{M}$ is teleported to the outside due to the known projection. After the black hole evaporates, the radiation becomes a pure state. In the HM proposal, the information loss is circumvented by requiring a unique result of the measurement and the probability of the postselected state is removed by renormalizing the state. However, such deterministic projection is not a legitimate operation in the framework of quantum mechanics. 

In some sense, to recover the unitarity outside the black hole, the nonunitary operation is involved near the singularity in the HM model. However, for HM model to conserve information there are more restrictions. In the Comment \cite{Gottesman:2003up}, Gottesman and Preskill reasoned that since the infalling radiation interacts with the black hole as it falls to the singularity, there is a global evolution acting on $\mathcal H_\text{M}\otimes \mathcal H_\text{in}$. Then the boundary state $|BH\rangle_{\text{M}\otimes \text{in}}$ has to be changed accordingly, as suggested by Eq. (\ref{eq:U_AA}). This looks like a conspiracy. In short, very particular final states and interactions have to be chosen to achieve the claimed result. Otherwise, the information loss is almost unavoidable in the HM model \cite{Lee:2020aft,Gottesman:2003up}. The model in our study is free from this ailment. The measurement occurs at the island which appears near the horizon after Page time. The infalling Hawking particles are measured immediately after being created at the horizon. The evolution due to their interaction with the black hole during the journey to the singularity no longer exists. Therefore, it is free of such conspiracy. The initial period of the evaporation is the same as Hawking predicted until Page time when a transition occurs to the wormhole saddle with openings at the horizon. 

The HM proposal is believed to reconcile the smoothness of the horizon with unitarity \cite{Lloyd:2013bza} but we believe that this viewpoint is not without question. First of all, the projection is not a unitary operation even if we post-select the state. As remarked in \cite{Gottesman:2003up}, such measurement with pre-determined results is at odds with quantum mechanics and violates causality. By renormalizing the state, the final-state projection is more properly understood as the boundary condition in the future which we can unitarily evolve backwards to obtain all the information. However, such boundary condition cannot be realized from the initial Unruh vacuum if we assume locality and unitarity. A simple argument is the following. Given a Hawking pair is formed near the horizon,
\begin{equation}
    |BH\rangle\otimes|R\rangle_{\text{in}\otimes\text{out}} = \frac 1 {\sqrt{N}} \sum_j |BH\rangle\otimes |j\rangle_\text{in}\otimes |j\rangle_\text{out}\,.
\end{equation}
Local unitary operations inside the black hole can mix the infalling states and the black hole states as follows, 
\begin{gather}
    |BH\rangle\otimes|R\rangle_{\text{in}\otimes\text{out}}\rightarrow \frac 1 {\sqrt{N}} \sum_{j} (U_{BH\otimes in}\otimes I)|BH\rangle\otimes |j\rangle_\text{in}\otimes |j\rangle_\text{out}\,,
\end{gather}
where $U$ is the unitary operation on the black hole state $|BH\rangle$ and the infalling state $|j\rangle_{in}$. This operator does not change the entanglement between the outgoing states $|j\rangle_{\rm out}$ and the total state inside the black hole $\sum_{j} (U\otimes I)|BH\rangle\otimes |j\rangle_\text{in}$ since the density matrix of the outgoing qubit is invariant under such unitary operations
\begin{gather}
    \rho_{\rm out}=tr_{BH\otimes in} |BH\rangle\otimes|R\rangle_{\text{in}\otimes\text{out}} \langle R|\otimes\langle BH|\rightarrow \rho_{\rm out}\,.
\end{gather}
The fine-grained entropy inside the black hole does not change and therefore the initial state cannot be in the assumed state. 

In our island-measurement correspondence model, the black hole is described as an ordinary quantum system seen from the outside, no operations that are forbidden by quantum mechanics are allowed. The horizon is smooth in this proposal since measurements do not retro-influence the Unruh state at the horizon. Therefore, it is free of firewalls. Furthermore, the HM model was proposed before the discovery of gravitational entropy formula and the entanglement wedge reconstruction does not have a correspondence in the model. For example, the information about states in the entanglement wedge cannot be known from the radiation except the black hole states. This is taken into account in this study. The information about the black hole can be viewed as on the surface of the horizon. Initially, no information can be drawn from the black hole since the Hawking radiation is thermal, or equivalently, the entanglement wedge of the radiation only includes regions away from the black hole. After Page time, the states entering the island are measured and the remaining states correspond bijectively with the states in the radiation according to the entanglement swapping protocol.

The appeal of the HM proposal is that the collapse of the wavefunction, or the measurement process occurs at the singularity instead of at the horizon where semiclassical descriptions are expected to apply if issues such as the trans-Planckian problem are ignored. As the state at the singularity is fixed, the infalling radiations must be subjected to the boundary condition throughout the history of the black hole from the very beginning. This gives a continuous information leaking picture which is different from what Page argued \cite{Page:1993wv,Page:1993df}. The consequence of it is a constant trivial Page curve---the entropy of the black hole or the radiation is always zero. In other words, the black hole just {\it cannot} entangle with anything!\footnote{To be precise, it can entangle only for a very brief time. Then the entanglement has to be broken and the black hole information is teleported out.}  

The fast scrambling of black hole along with the bulk reconstruction all point to the direction that such final state projection, if exists, already occurs when information enters the island near the horizon instead of waiting until it hits the singularity. This is also suggested by the bulk reconstruction from AdS/CFT that the states in the entanglement wedge can be retrieved from the boundary operators. This is all on the assumption of a unitary evaporation. We argue that it is more natural to introduce a measurement process than a post-selected measurement process. Then paradox of the information loss or conservation can be directly attributed to the same problem of unitarity and locality about measurements. This connection may suggest that the Page transition can be understood as a quantum-to-classical transition at which the quantum states are effectively measured and wavefunctions are spontaneously collapsed. The issues regarding quantum measurement and its interpretations are beyond the scope of this study \cite{bassi2013models,schlosshauer2005decoherence,bong2020strong}.

\subsection{Osuga-Page firewall-free qubit model} \label{Subsec:OP}
Osuga and Page considered a qubit model which allows information to escape and was argued to be free of firewalls \cite{Osuga:2016htn}. In their setup, the non-local gravitational degrees of freedom are assumed. The gist of their consideration is the following. The maximally entangled radiation qubits are created near the horizon in an ordinary Hawking process. As the outgoing qubit, representing the radiation, propagates outward from the horizon, its interaction with the nonlocal gravitational qubit induces a unitary transformation which turns the mixed state outgoing qubit into a pure state. An example of such the evolution operator is $e^{-i\pi(1-K/K_h)P_{ac}}$ such that it obeys 
\begin{gather}
    \lim_{r\rightarrow \infty} e^{-i\pi(1-K/K_h)P_{ac}}|q\rangle_a |B\rangle_{bc}=-|B\rangle_{ab}|q\rangle_c\,,
\end{gather}
where $|B\rangle_{ab}$ defined by
\begin{align}
    |B\rangle_{ab}=\frac{1}{\sqrt{2}}(|0\rangle_a|1\rangle_b-|1\rangle_a|0\rangle_b)
\end{align}
is the singlet Bell state. $P_{ac}=|B\rangle_{ac}\langle B|$ is the projection operator, and $K$ is some curvature invariant such as the Kretschmann scalar which decays from $K_h$ at horizon to zero at infinity. Here $a$ represents the black hole nonlocal qubits, $b$ is the infalling qubits and $c$ is the outgoing qubits.


Different from the usual wisdom of the nonlocal interactions between local qubits for resolving information issue of the black holes, the OP model suggests that information can come out due to the interactions between the Hawking radiation and the nonlocal black hole qubits. As the radiation propogates to the infinity, an teleportation-like process is effectively realized which switches the radiation qubits with the black hole qubits. However, this argument seems to be problematic if one collects the radiation at a finite distance from the black hole. When the radiations are collected at any positions with nonvanishing $K$, they will be found in a mixed state according to this prescription, as a consequence, unitarity and entropy will not preserved. On the other hand, even if the radiations are allowed to propagate to the infinity, the OP model appears to predict a trivial Page curve which is at odds with the recent results from the GPI. Such realization is similar to assuming that the effective measurements are nonlocal unitary operations in our proposal. Under this assumption, the teleportation protocol can be completed and the information is preserved. 



\subsection{Connection with Reeh-Schlieder theorem}\label{Subsec:RS}
It is widely believed that once the infalling matter has passed through the event horizon, it can no longer influence the outside due to causality \cite{Perry:2021mch}. It is also assumed that either quantum mechanics breaks down or some new physics is required in order for the information inside the horizon to escape. However, axioms of quantum mechanics not only have the unitary evolution but also includes processes as measurements which may not be naively understood as just another unitary operation, otherwise teleportation will not be realized. The Reeh-Schlieder (RS) theorem states that the vacuum is a cyclic separating vector of the algebra of operators in any open region $\mathcal{U}$ \cite{Reeh:1961ujh}. It means that a dense set of states in the Hilbert space $\mathcal{H}$ can be produced by local operators acting on the vacuum. Intuitively, it suggest that one can ``create the Moon" from local operations inside the house. The requirement is that such operations have to be nonunitary and the vacuum is fully entangled.\footnote{The set of local Hermitian operators are also capable of such Moon construction.} For a nice review on RS theorem, see Ref.~\cite{Witten:2018zxz}. For a finite dimensional system, the RS theorem implies that for two maximally entangled subsystems that are spacelike separated, we can apply local operations like projections to manipulate the partner entangled subsystem. The set of such manipulations is dense in the space of operations. The quantum teleportation and entanglement swapping are examples of such operation provided that the measurement operator is nonunitary. For teleportation and entanglement swapping, classical communications are needed in order to have a complete control of what the states one can build at the other end since the results of the measurements are not known beforehand unlike the predetermined projections. If the measurements are some local unitary operators, such operation will not be possible. The RS theorem suggests a possible resolution of the seemingly acausal information transmission of black holes by nonunitary operators inside the black hole, which is similar to what we propose in this paper.

\subsection{Retrospect on no-cloning theorem in black holes}\label{Subsec:no-cloning}

\begin{figure}[ht]
    \centering
    \includegraphics[width=0.6\textwidth]{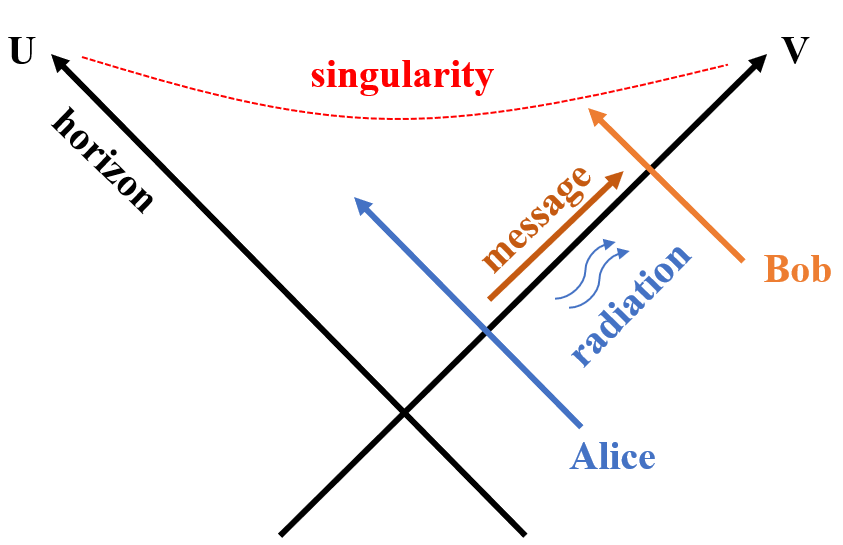}
    \caption{The Alice-Bob test of quantum no-cloning theorem in the black hole.}
    \label{Fig:nocloning}
\end{figure}

It was proposed by 't Hooft et al. that the information about the black hole interior can be equivalently described by the data on the surface of the horizon \cite{tHooft:1993dmi,Susskind:1994vu}. The apparent paradox of the quantum cloning in some Cauchy surfaces across the horizon was shown to be avoided since no time-like observer is able to access both the ``cloned state" and the original state without hitting the singularity beforehand. The argument provided in \cite{Maldacena:2013xja} has a drawback of assuming that the information is sent into the black hole before the Page time. In Ref.~\cite{Hayden:2007cs}, the quantum state is assumed to be thrown into the old black hole after Page time, and it is shown that the information is ``reflected" back from the horizon almost immediately. The diagrammatic illustration is shown in Fig.~\ref{Fig:nocloning}. In below, we briefly review their argument. 

The Kruskal coordinates are related to the Schwarzchild coordinates by the following transformations, 
\begin{gather}
    U=-e^{(r_\ast-t)/2R_s},\quad V=e^{(r_\ast+t)/2R_s}, \quad {\rm where} \quad r_\ast=r+R_s\ln{(r-R_s)/R_s}\,,
\end{gather}
and the singularity lies at $UV=1$. Alice has to send off the state before Bob hits the singularity at $U_A=V_B^{-1}$. The proper time that Alice has to communicate with Bob is 
\begin{gather}
    \tau_A\propto R_s(V_A/V_B)=R_s e^{-\Delta t/2R_s}\,,
\end{gather}
where the proportional constant is order one if Alice begins the free fall from rest outside the horizon, $V_A$ is the time when Alice crosses the horizon and $\Delta  t$ is the time Bob delays with respect to Alice. The shortest time to retrieve the information is given by the scrambling time assuming the random unitaries inside the black hole, otherwise Bob may be able to jump in with Alice's information even earlier than the scambling time. The time that Alice takes to reach the island is around $2R_s\ln(S_{\rm BH}/N)$, where $N$ is the CFT central charge. This is the same as the scrambling time given in \cite{Hayden:2007cs,Sekino:2008he}. Assuming that Bob receives the information after the scrambling time and jumps into the black hole immediately after receiving the information, we obtain the maximal proper time $\tau_{A,max}\approx N/R_s$ that Alice has to send her message before Bob hits the singularity. Therefore, in order to send the information to Bob, the signal has to be super-Planckian. 


The above argument shows that although we can draw a spacial slice with two copies of the information, no time-like or null observer is able to access both copies. This is sometimes referred to as the black hole complementarity \cite{Susskind:1993if,tHooft:1990fkf}. The essential statement is that there is dual description of the information falling into the black hole --- for an outside observer the infalling objects take infinite time to approach the horizon and thus the information is stored on the surface of the horizon; for an infalling observer, the horizon is not a detectable object and information can go through it smoothly to the inside of the black hole. The linearity of quantum mechanics is still violated but since no one is able to detect the violation, this problem is swept under the rug. To preserve unitarity and the smoothness of the horizon simultaneously, such argument seems unavoidable. On the other hand, the no-cloning problem can be avoided by a energetically singular horizon such as a firewall, or in our case, an informationally singular ``purification wall" after Page time. By introducing the island--measurement correspondence, the teleportation protocol can only purify the states outside but is not able to transfer the information due to a causal horizon. The radiation is purified into a random ensemble and therefore the violation of no-cloning theorem is avoided. In this case, the no-cloning theorem can be preserved without entailing the practical detectablility of the violation. If we adopt a different interpretation of the measurements, assuming that one has a complete knowledge about the black hole forming history as well as the interactions inside and can predict all the ``measurement" outcomes, then the teleportation protocol can be completed and the information will be teleported out from the black hole in terms of radiation. In this case, the violation of the no-cloning theorem is inevitable unless all infalling information is teleported out exactly at the horizon or slightly outside (e.g. islands in eternal black holes). But such realization requires that the measurement processes are completed in Planck time.

\section{Regarding the Page curves in 2D asymptotically-flat gravity}\label{Sec:2Dmodel}
The islands and Page curves of various 2D evaporating black holes have been explicitly studied in Refs.~\cite{Wang:2021mqq,Hartman:2020swn,Gautason:2020tmk}. The issue concerning the different rates of entropy change in the initial and final stage of the Page curve, which may not be simply understood as an artifact of nonequilibrium thermodynamics was brought up in \cite{Wang:2021mqq}. 
We assume that regardless of the details of quantum gravity in the strongly gravitating region, far from the black hole in an environment like we are now living in, one has a clear quantum mechanical description of the radiation states and the black hole states. We show that the informational interpretation of the islands proposed in this paper eases the tension between the different purification rates and reproduces the right Page curves for the 2D asymptotically flat gravities.

\begin{figure}[ht]
    \centering
    \includegraphics[width=0.56\textwidth]{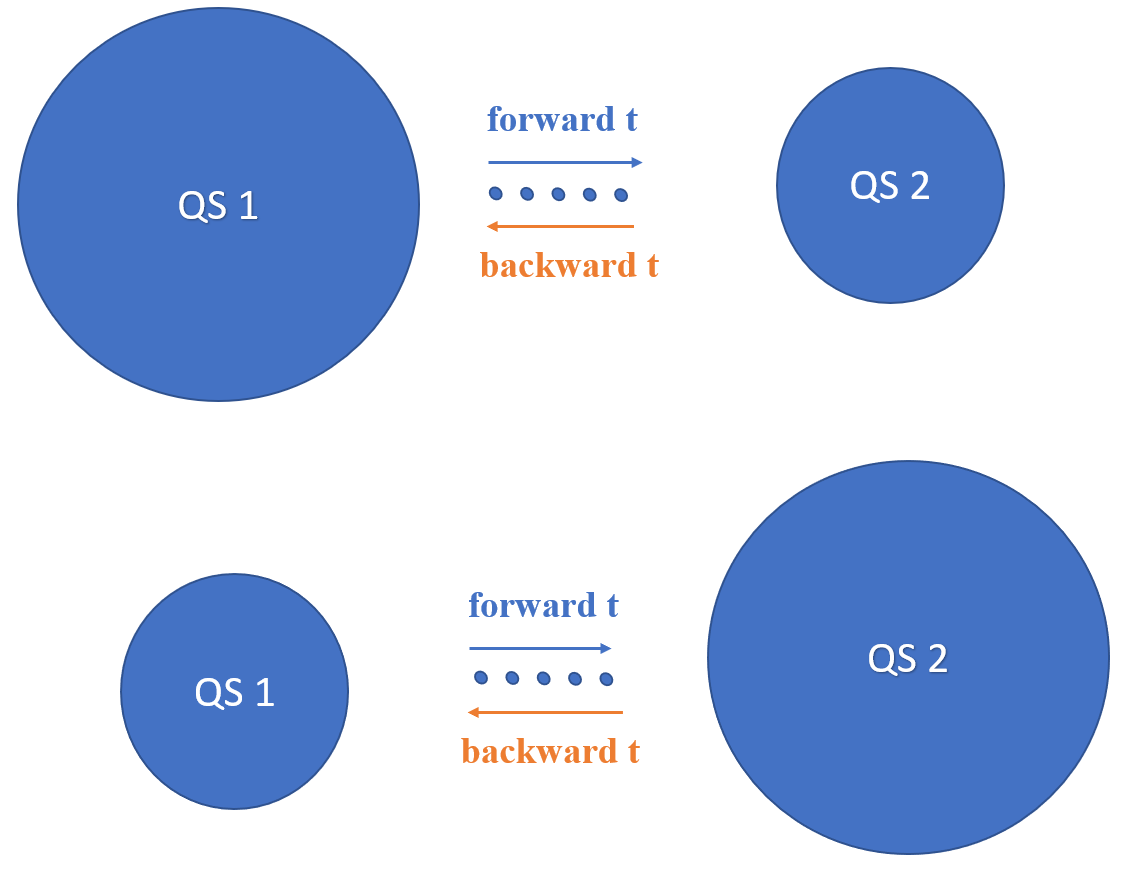}
    \caption{Assuming that the black hole is an ordinary quantum system, we can apply the Page theorem at the initial and final stages of the evaporation. The upper diagram represents the initial stage of evaporation where the black hole (QS 1) is viewed as the larger system. The lower diagram represents the late stage of evaporation where the radiation (QS 2) has larger degrees of freedom. The upper diagram and the lower graph are indistinguishable under time reversal if the evolution is unitary. Given that the black hole has a constant rate of evaporation, one should expect a symmetric Page curve as in \cite{Page:1993wv}.}
    \label{Fig:a}
\end{figure}

\begin{figure}[ht]
    \centering
    \includegraphics[width=0.7\textwidth]{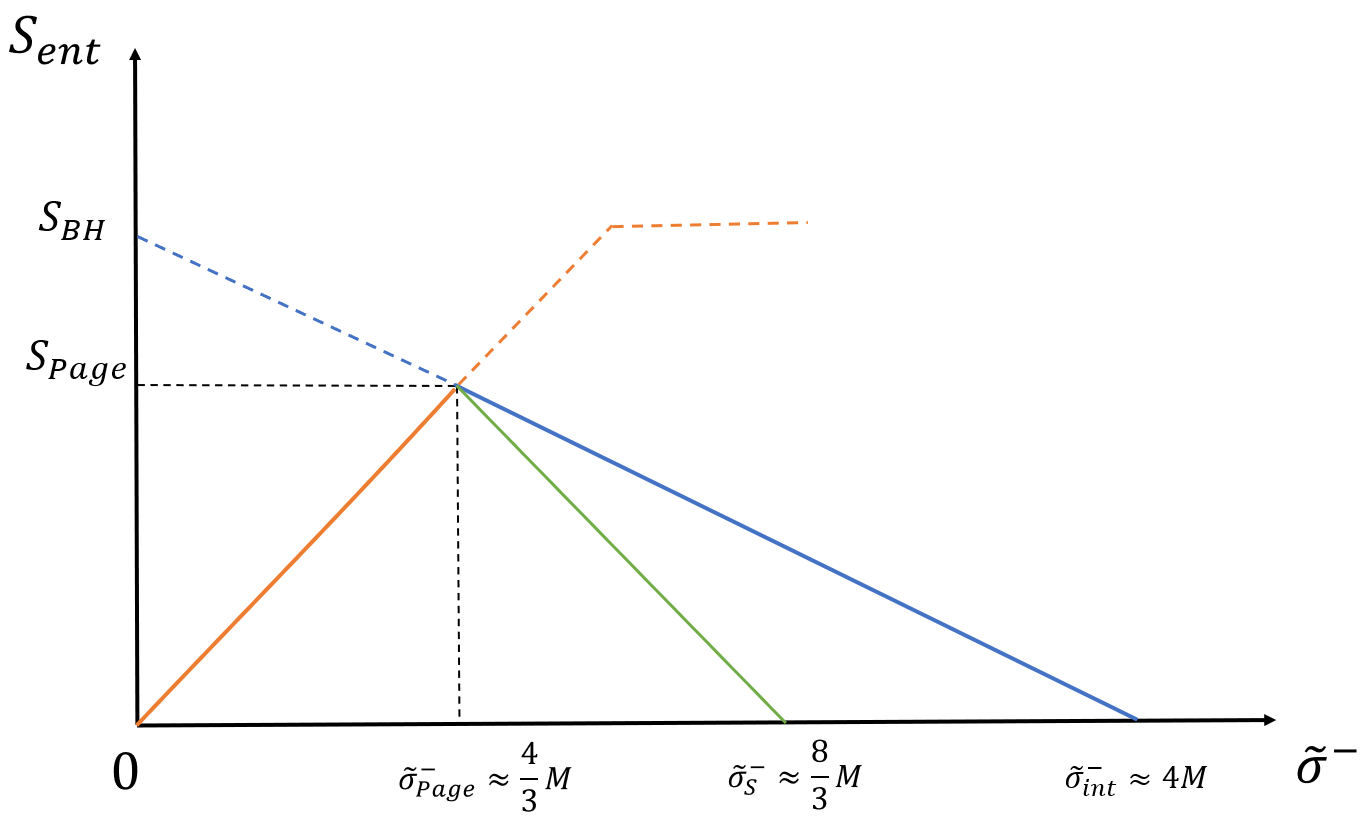}
    \caption{Page curves for 2D asymptotically flat BPP-RST models. $S_{\rm BH}$ is the Bekenstein-Hawking entropy of the initial black hole with ADM mass $M$. Page transition occurs at time $\tilde\sigma^-_{Page}=\frac{4}{3}M$. The solid blue line is the Page curve calculated from the island formula. The green line is from the Page's theorem which is symmetric to the orange line.}
    \label{Fig:b}
\end{figure}

The gist of the arguments is summarized as follows. Assuming the validity of Page's theorem which treats the black hole and its radiation as two quantum systems, the radiation processes in the initial and final stages are equivalent up to a different identifications of the large and small systems. Therefore, the initial period leads to entanglement generation and the final period purification. Given the random Haar unitary, the emitted radiation at the beginning of the evaporation is maximally entangled with the black hole. This is due to that the black hole has larger degrees of freedom than the total radiation emitted [see Fig.~\ref{Fig:a}]. Consequently, the new emitted radiation in a short time interval is entangled with the large system, the black hole. This results to the increase in the entropy. In the final process, the roles of the black hole and the radiation switch and the total radiation is the larger system. In this case, the new emitted radiation is maximally entangled with the early radiation. This process is the purification of early radiation states which results in entropy reduction. Since the Hawking radiation is at a constant rate with temperature $T_{\rm BH}=\frac{1}{2\pi}$ for the 2D dilaton gravity, the rate of entropy changes should be the same up to a minus sign according to this argument. There is a more succinct way of understanding the above argument. The evaporation process is assumed to be unitary, then one can simply reverse the time and run the evaporation backwards. The initial stage in the forward time and the final stage in the backward evolution are identical if we only keep track of the sizes of the systems.\footnote{One can argue that the radiation process is irreversible and the time reversal symmetry is not valid. However, the irreversiblity is in the thermodynamic sense and applies to the coarse-grained quantities which is not relevant here. The entanglement entropy, on the other hand, is always the same for both the radiation and the black hole whether or not they are in equilibrium. Furthermore, the thermodynamic entropy of the photon gas is dependent on quantities such as the spectrum density and the size of the container, and may not be naively identified with the fine-grained entropy when a nonequilibrium process is involved.} This argument using Page's theorem in quantum informational picture produces a symmetrical Page curve which contradicts the Page curve calculated from the quantum RT formula. From the quantum RT prescription, the entanglement entropy of the black hole is 
\begin{equation}
 S_{\rm ent}  =  \min \left\{\mathrm{ext}_I\left[ \frac{\mathrm{Area}(\partial I)}{4G_{\rm N}}  + S_{\rm matter}(R\cup I) \right]\right\} \,,
\end{equation}
where $I$ represents the entanglement island. Roughly speaking, before Page time the RT surface is trivial, the entropy is given by the semi-classical calculation
\begin{equation}
 S_{\rm ent} = S_{\rm matter}(R) \,,
\end{equation}
which increases linearly with time. After Page time, the RT surface lies at the horizon of the black hole and is the dominant term. Therefore
\begin{equation}
 S_{\rm ent}  \simeq  \frac{\mathrm{Area}(\partial I)}{4G_{\rm N}}  \,,
\end{equation}
which returns to the Bekenstein-Hawking entropy as Area($\partial I$) is the area of the horizon. The exact calculations give the Page curve as shown in Fig.~\ref{Fig:b}.

In the 2D dilaton gravity, the Bekenstein-Hawking entropy is given by $S_{\rm BH}=\frac{N}{6}M$, where $N$ is the number of scalar fields or the central charge of the CFT. In this case, the mass of the black hole is proportional to the area of the horizon, which counts the number of qubits in the unit of Planck area. For an evaporating black hole, island appears when the black hole has radiated a third of its mass. At this transition point, the remaining black hole mass $\frac{2}{3}M$ is twice of that radiated out $\frac{1}{3}M$. We ignore the issues on the state counting of the Hilbert spaces of black holes which can be explained by nonorthogonality of semiclassical states. Adopting the quantum mechanical description where we imagine that we can only access to the region slightly outside the horizon, therefore we cannot be sure whether the inside is a black hole or the black hole has completely evaporated into a thermal gas. Since local unitary operations inside the horizon do not affect the entanglement structure between the black hole and the radiation. For simplicity, we can pick some unitary operator $\mathcal{U}_{BH}$ on black hole states such that the post-operation system has the following property --- a half of its qubits are pure states and the other half are maximally entangled with the radiation. For simplicity, we first consider the situation of no interactions between the black hole states. The island-measurement correspondence tells us that in the statistical sense, the teleportation and the entanglement swapping protocols are implemented with equal probability (the final step of the classical communication is not pertinent to the entanglement entropy). Therefore, half of the outgoing radiations are pure states and half are entangled with early radiations statistically. This gives the expected purification rate calculated from the fine-grained entropy formula [the blue curve in Fig.\ref{Fig:b}]. Biases from this ratio will be automatically corrected in the evaporation and therefore the entropy has a stable decline at half the rate of the initial increase. 

\begin{figure}[ht]
    \centering
    \includegraphics[width=\textwidth]{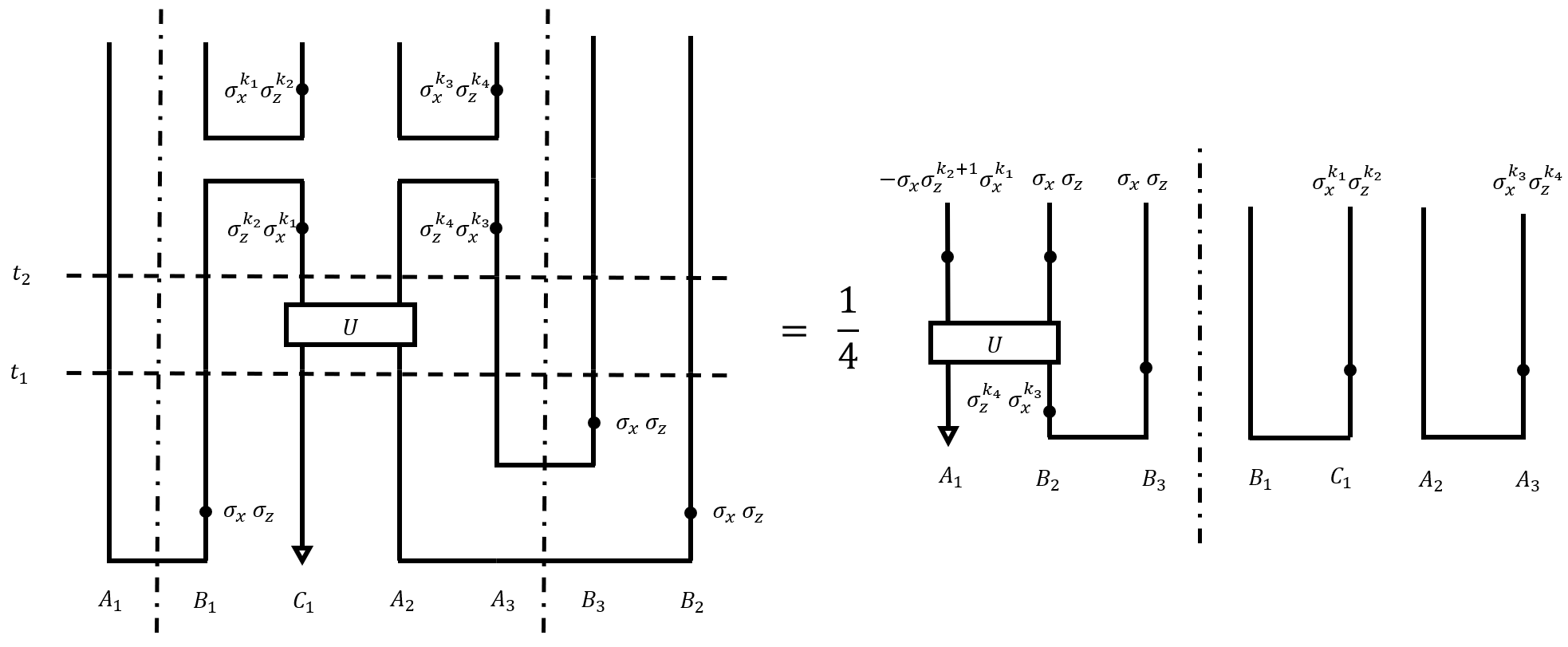}
    \caption{Mixing of teleportation and entanglement swapping. The time direction goes up. On the LHS, below $t_1$ is the initial state. From $t_1$ to $t_2$, the early infalling radiation $\rho_{A_2}$ evolves unitarity with the initial black hole pure state $|\psi\rangle_{C_1}$. Above $t_2$, the Bell states projections are applied. The RHS is the final state. The region between the two vertical lines for the left diagram are inside the horizon. For the right diagram, the region inside the horizon is to the right of the vertical line.}
    \label{Fig:c}
\end{figure}

Considering the unitary evolution between the early infalling state and the initial black hole state, one gets a mixed teleportation and entanglement swapping when the measurements are performed. To be concrete, we can consider a micro-process where one such measurement is performed. Suppose that the pure state $|\psi\rangle_{C_1}$ represents the black hole initial state. The singlet state $|\Psi^-\rangle_{A_2B_2}$ represents the entangled state of early Hawking particle and the black hole state, where $A_2$ denotes the state inside the horizon. We introduce the scrambling evolution $U$ inside the black hole acting on $|\psi\rangle_{C_1}\langle\psi|\otimes \rho_{A_2}$. After the page time, the island is formed and Bell measurements are evoked. Suppose that we have the new formed singlet states $|\Psi^-\rangle_{A_1B_1}$ and $|\Psi^-\rangle_{A_3B_3}$, where $B_1$ and $A_3$ denote the state inside the horizon (the region between the two dashed vertical lines in Fig.~\ref{Fig:c}). Bell measurements are performed either on the recent infalling radiation state and the black hole state (teleportation) or on the recent and the early infalling radiation state (entanglement swapping). Combining Eqs.~(\ref{eq:tele_eq}) and (\ref{eq:es_eq}), we have the teleportation equation for the mixed process,
\begin{multline}
\label{eq:tele3}
\left(1\!\!1_{A_1B_1}\otimes U_{C_1A_2}\otimes 1\!\!1_{B_2A_3B_3}\right)|\Psi^-\rangle_{A_1B_1}\otimes |\psi\rangle_{C_1}\otimes |\Psi^-\rangle_{A_2B_2}\otimes |\Psi^-\rangle_{A_3B_3} \\
= \frac 1 4\sum_{i=1}^4\sum_{k_i=0}^1 \tilde U(k_1,k_2,k_3,k_4)_{A_1B_2} |\psi\rangle_{A_1}\otimes |\Psi^-\rangle_{B_2B_3}\\
\otimes \left(1\!\!1_2\otimes \sigma_x^{k_1}\sigma_z^{k_2}\right)|\Phi^+\rangle_{B_1C_1}\otimes \left(1\!\!1_2\otimes \sigma_x^{k_3}\sigma_z^{k_4}\right)|\Phi^+\rangle_{A_2A_3}\,,
\end{multline}
where 
$$
\tilde U(k_1,k_2,k_3,k_4) = -\left(\sigma_x\sigma_z^{k_2+1}\sigma_x^{k_1}\otimes \sigma_x\sigma_z\right)U(1\!\!1_2\otimes \sigma_z^{k_4}\sigma_x^{k_3})\,.
$$
We can see that the scrambling evolution $U$ up to some Pauli matrices is also teleported out along with the states. The evolution $U$ can be decoded from outside if the measurement results $(k_1,k_2,k_3,k_4)$ are known. Therefore, such interaction processes will not influence the validity of the results of the simple analysis given in the last paragraph. Therefore, such mechanism reproduces the Page curve in Fig.~\ref{Fig:b}. The diagrammatic representation of this algebra is shown in Fig.~\ref{Fig:c} which we will describe in detail below.

Above description is similar to the concept of teleportation-based quantum computation \cite{Gottesman99}, in which there are some single- or two-qubit gates applied before the teleportation or entanglement swapping processes. The diagrammatic representation of teleportation-based quantum computation has been studied before in a different context \cite{Zhang13,Zhang16}. The mixing of teleportation and entanglement swapping process can also be presented diagrammatically as shown in Fig.~\ref{Fig:c}. The Bell states $|\Psi\rangle$ are represented by a cup $\bigsqcup$. Correspondingly, the projection on Bell states are represented by joining a top cup $\sqcup$ and a bottom cap $\sqcap$. Local Pauli matrices are represented by the single dots, which can be moved along the connected lines. On the left, we have the black hole unentangled state $C_1$, the entangled state $A_2$ with the early radiation $B_2$. $A_{1,3}$ and $B_{1,3}$ are the new Hawking pairs generated near the horizon. After the unitary operation $U$, the end state (on the top) is a pure state $A_1$ outside the horizon, two entangled pairs inside the horizon and an entangled pair outiside horizon $B_{2,3}$. The connected lines can be straightened, corresponding to the information flow. The bits $k_1,k_2,k_3,k_4\in\{0,1\}$ correspond to the random measurement results. If final state projection has been applied, then those bits are fixed as 1. The LHS of Fig. \ref{Fig:c} can be deformed into the RHS, where the factor $1/4$ is required for normalization. The diagram on the right shows that the left diagram is equivalent to a unitary operation outside the horizon with a pure state and a pair of maximally entangled state outside. Fig. \ref{Fig:c} is equivalent to the teleportation equation given by Eq. (\ref{eq:tele3}). Such diagrammatical rules are based on the Temperley-Lieb algebra \cite{Temperley71}, in which Bell states give a representation on $\mathcal H_2\otimes\mathcal H_2$.

In addition, for eternal black holes in Hartle-Hawking states, the fine-grained entropy saturates at the Bekenstein-Hawking value $S_{\rm BH}$. The infalling thermal gas into the black hole, which is assumed to be in the thermal field double state, adds to the entanglement entropy of the black hole. Since the entanglement entropy of the black hole is fixed, the radiation process has to reduce the entanglement entropy of the black hole equally fast. This is consistent with the entanglement swapping model discussed in this paper.

\section{Discussion}\label{Sec:discussion}
The logic for this proposal is simple. If the energetically singular horizon has to be avoided, the initial Unruh state of the radiation needs to be valid. This initial radiation state, if propagates to the null infinity without any nonlocal effect, does not conserve the fine-grained entropy or carry any information. If the fine-grained entropy is required to return to its initial value (zero for a pure state black hole), then the entanglement between the late Hawking radiation and the black hole has to be cut off after the creation of the particle to purify the early radiation. If the entanglement is cut off after the creation, the operation cannot be described by a local unitary operator. Within the standard quantum mechanics, there exists but one such operation---the measurement, that effectively occurs inside the black hole and purifies the early Hawking radiation. Whether or not the measurement can be treated as an emergent operator from a soup of complex processes, its interpretation has been a conundrum. If the additional information conservation is required, the ``measurement" results cannot be completely random. If the interpretation of bulk reconstruction is accurate, such measurement processes have to occur at the surface of the island boundary which lies at the horizon up to Planck length. The exact position of the horizon is not well-defined quantum mechanically, so we do not differentiate the two after Page time in our consideration. Through the quantum mechanical view, the issues of nonlocality and nonunitarity in the measurement problem are identified with similar issues in the black hole situation even before considering the quantum gravity effects. A quantum mechanical view of the current understanding of black hole informational paradox is what we aim for in this paper. 

One should note that the cross-lightcone state manipulation (or quantum steering) is common in quantum information theory. Acausal information transfer would be possible if measurement results are not quantum random, such as the final-state projection. Since we only have measurements with uncontrollable results as the potential nonunitary operators within the basic frame of quantum mechanics and thus no such effective information transmission can be realized. This is based on the assumption that the experimental results are fundamentally unpredictable even if we have all the initial data about the apparatus and the interactions. Similarly, the information inside the black hole is causally segregated by the horizon from the states outside. As required by unitarity, the information inside the black hole must somehow escape to the outside asymptotic region through a Hawking channel which does not carry initial data when the radiation leaves the horizon. In a standard quantum measurement of two entangled particles (A and B), $|\psi\rangle_B$ can change ``acausally" from a mixed state to a pure state when a local measurement is conducted on A regardless of whether or not the two events are within the light-cones of one another. The result is an ensemble of pure states. Similar nonunitary processes inside the black hole also purify the states outside viewing from outside. Similar phenomenon through nonunitary operations on vacuum is demonstrated by the Reeh-Schlieder theorem in quantum field theory. Therefore, if we require that the outside radiation is purified and that the initial Hawking particles do not carry information, then nonunitary operators are unavoidable inside the black hole. If we further require that the information inside the black hole comes out across the causal horizon (or unitarity in the entire process of evaporation), then viewing from the outside a deterministic measurement result is necessary once one gains a complete knowledge about the state of the black hole and the interactions. In this case, the measurement is a nonlocal but unitary operation. On the other hand, if measurements in quantum mechanics are fundamentally nonunitary processes, it should not be too surprising that information can be lost in the black hole. We reemphasize that although the two scenarios give the same Page curve, they do not all guarantee the conservation of information which depends on interpretations of measurements.

The island--measurement proposal is equivalent to saying that measurement-like processes can occur spontaneously or effectively without involving humans or apparatus, and thus macroscopic objects do not have to evolve completely unitarily once the classical transition occurs. The current status on the understanding of measurements is tumultuous and confusing. There are as many different proposals as the resolutions of black hole information puzzle. Adopting the nonlocal unitary interpretation in the measurement theory grants similar processes in the black hole scenario without recourse to a nonlocal quantum gravity theory. If one accepts the standard interpretation of measurements as objective wavefunction collapse, such measurement(-like) effects can greatly influence our understanding of macroscopic objects like black holes. One can imagine two microscopically identical black holes, one formed from pure state matter and the other from maximally mixed matter entangled with a controlled sample. A scrambling time after the formation, we will observe that one black hole emits the expected thermal quanta entangled with it and the other emits quanta which are entangled with the controlled sample. Therefore, the state of the radiation which is derived from the spacetime geometry is determined by the purity of the black hole. Assuming a smooth horizon leads to the conclusion that a sufficiently mixed black hole will purify the Hawking radiation through certain nonlocal effect and we provide one mechanism in the context of standard quantum information theory. If this conclusion can be extended beyond the background of this study, it may point to the quantum-to-classical transition which is determined by the purity of states, especially for the strong gravity systems. This is different from the gravitational wavefunction collapse or Di\'osi–Penrose model discussed in refs.~\cite{penrose1996gravity,Diosi:1986nu,Diosi:1988uy,Donadi:2020kzc} where the authors argued that gravitational potentials can destroy the superposition of quantum states. 

In this proposal, there are no firewalls. The Unruh vacuum remains intact so that a freely falling observer does not experience the energetic curtain at the horizon. Instead, the observer is measured against the infalling states and is potentially teleported out in terms of Hawking radiation. This scenario preserves the smooth horizon but does not offer a save boat for an infalling observer to cross the horizon. There are no fundamental limitations on teleportation and entanglement swapping applied to macroscopic and living objects except for the technology to process huge amounts of data.

One way to go around the possibility discussed in the paper is that the information comes out from the black hole due to the nonperturbative corrections. It is questionable whether a nonperturbative construction of gravity can resolve issues around black hole radiation. As we can imagine an arbitrarily large black hole such that the gravity at the horizon is arbitrarily weak. In the arbitrarily weak gravity, we can safely apply EFT and the issues are the same without any amelioration. Many studies suggest a resolution from the overcomplete basis of semi-classical states. This idea is also applied to resolve the state-counting issue in the black holes \cite{Hsin:2020mfa,Stanford:2020wkf}. This idea is related to what we postulate in this study as certain form of ensemble averages of semi-random unitary theories or states is assumed to account for the nonvanishing off-diagonal terms in the double copies of density matrix, i.e. $|\langle \psi_i|\psi_j\rangle|^2$ but not in the single copy. Besides, studies of islands in eternal black holes which are solutions of Einstein's equations suggested certain topological changes based on the purity (or mixedness) of the black holes even if they have constant energies or masses and are in thermal equilibrium with radiations \cite{Wang:2021woy,Hashimoto:2020cas}. This is beyond the classical gravity theory which has no hair, and it is unclear if this is a sheer artifact of the entropy calculations. The exotic islands which are not at the black-hole horizons in certain specific geometries are not considered in this study \cite{Li:2021lfo,Sybesma:2020fxg}. Besides, one should note that there are many other completely different approaches to the information paradox. For a nonexhaustive list, the antipodal identification changes the horizon from a $\mathcal{S}^2$ to a $\mathcal{PS}^2$ to avoid the interior of the black hole \cite{tHooft:2018zwd}; the classical ``firewall" solution completely removes the trapped surface of black holes \cite{Kaplan:2018dqx,McManus:2020lgm}; the soft hair proposal requires black hole to carry a large amount of soft supertranslation modes \cite{Hawking:2016msc}. For a nice review, see ref.~\cite{Harlow:2014yka}.

In the end, it is not evident exactly when the information of the black hole state $|m\rangle$ is transferred to the outside of the black hole if certain boundary condition is given beforehand. In particular, recent experiments on delayed-choice entanglement swapping suggest that the purification may occur even before the projection is implemented \cite{ma2012experimental,Ma:2014dta}. This has to do with the basic understanding of wavefunctions. It suggests that even in the final-state projection model, it is questionable if high frequency excitations can be detected at the horizon by a freely falling observer. We may run into, at least, a ``not-so-smooth" horizon dilemma. However, the exact implications of these experiments are still not completely clear at the present and we will end our discussion here.

\acknowledgments
X.W. wants to thank Ran Li and Tim Hollowood for the questions and comments, also thank Watse Sybesma, Hong Wang and Di Wang for helpful discussions.



\providecommand{\href}[2]{#2}\begingroup\raggedright\endgroup


\begin{thebibliography}{10}

\bibitem{Giddings:2006sj}
S.~B. Giddings, {\it {Black hole information, unitarity, and nonlocality}},
  {\em Phys. Rev. D} {\bf 74} (2006) 106005,
  [\href{http://arxiv.org/abs/hep-th/0605196}{{\tt hep-th/0605196}}].

\bibitem{Banks:2010zn}
T.~Banks and N.~Seiberg, {\it {Symmetries and Strings in Field Theory and
  Gravity}},  {\em Phys. Rev. D} {\bf 83} (2011) 084019,
  [\href{http://arxiv.org/abs/1011.5120}{{\tt arXiv:1011.5120}}].

\bibitem{Bousso:2019ykv}
R.~Bousso and M.~Toma\v{s}evi\'c, {\it {Unitarity From a Smooth Horizon?}},
  {\em Phys. Rev. D} {\bf 102} (2020), no.~10 106019,
  [\href{http://arxiv.org/abs/1911.06305}{{\tt arXiv:1911.06305}}].

\bibitem{Gibbons:1976ue}
G.~W. Gibbons and S.~W. Hawking, {\it {Action Integrals and Partition Functions
  in Quantum Gravity}},  {\em Phys. Rev. D} {\bf 15} (1977) 2752--2756.

\bibitem{Hawking:1978jz}
S.~W. Hawking, {\it {Quantum Gravity and Path Integrals}},  {\em Phys. Rev. D}
  {\bf 18} (1978) 1747--1753.

\bibitem{Marolf:2020rpm}
D.~Marolf and H.~Maxfield, {\it {Observations of Hawking radiation: the Page
  curve and baby universes}},  {\em JHEP} {\bf 04} (2021) 272,
  [\href{http://arxiv.org/abs/2010.06602}{{\tt arXiv:2010.06602}}].

\bibitem{Horowitz:2003he}
G.~T. Horowitz and J.~M. Maldacena, {\it {The Black hole final state}},  {\em
  JHEP} {\bf 02} (2004) 008, [\href{http://arxiv.org/abs/hep-th/0310281}{{\tt
  hep-th/0310281}}].

\bibitem{Gottesman:2003up}
D.~Gottesman and J.~Preskill, {\it {Comment on `The Black hole final state'}},
  {\em JHEP} {\bf 03} (2004) 026,
  [\href{http://arxiv.org/abs/hep-th/0311269}{{\tt hep-th/0311269}}].

\bibitem{Harlow:2014yoa}
D.~Harlow, {\it {Aspects of the Papadodimas-Raju Proposal for the Black Hole
  Interior}},  {\em JHEP} {\bf 11} (2014) 055,
  [\href{http://arxiv.org/abs/1405.1995}{{\tt arXiv:1405.1995}}].

\bibitem{Lloyd06}
S.~Lloyd, {\it Almost certain escape from black holes in final state projection
  models},  {\em Physical Review Letters} {\bf 96} (2006), no.~6 061302.

\bibitem{Lloyd:2013bza}
S.~Lloyd and J.~Preskill, {\it {Unitarity of black hole evaporation in
  final-state projection models}},  {\em JHEP} {\bf 08} (2014) 126,
  [\href{http://arxiv.org/abs/1308.4209}{{\tt arXiv:1308.4209}}].

\bibitem{Lloyd:2004wn}
S.~Lloyd, {\it {Almost certain escape from black holes}},  {\em Phys. Rev.
  Lett.} {\bf 96} (2006) 061302,
  [\href{http://arxiv.org/abs/quant-ph/0406205}{{\tt quant-ph/0406205}}].

\bibitem{Bousso:2013ifa}
R.~Bousso, {\it {Violations of the Equivalence Principle by a Nonlocally
  Reconstructed Vacuum at the Black Hole Horizon}},  {\em Phys. Rev. Lett.}
  {\bf 112} (2014), no.~4 041102, [\href{http://arxiv.org/abs/1308.3697}{{\tt
  arXiv:1308.3697}}].

\bibitem{Osuga:2016htn}
K.~Osuga and D.~N. Page, {\it {Qubit Transport Model for Unitary Black Hole
  Evaporation without Firewalls}},  {\em Phys. Rev. D} {\bf 97} (2018), no.~6
  066023, [\href{http://arxiv.org/abs/1607.04642}{{\tt arXiv:1607.04642}}].

\bibitem{Chatwin-Davies:2015hna}
A.~Chatwin-Davies, A.~S. Jermyn, and S.~M. Carroll, {\it {How to Recover a
  Qubit That Has Fallen Into a Black Hole}},  {\em Phys. Rev. Lett.} {\bf 115}
  (2015), no.~26 261302, [\href{http://arxiv.org/abs/1507.03592}{{\tt
  arXiv:1507.03592}}].

\bibitem{Borsten:2012fx}
L.~Borsten, M.~J. Duff, and P.~Levay, {\it {The black-hole/qubit
  correspondence: an up-to-date review}},  {\em Class. Quant. Grav.} {\bf 29}
  (2012) 224008, [\href{http://arxiv.org/abs/1206.3166}{{\tt
  arXiv:1206.3166}}].

\bibitem{Akil:2018iny}
A.~Akil, O.~Dahlsten, and L.~Modesto, {\it {Entanglement swapping in black
  holes: restoring predictability}},
  \href{http://arxiv.org/abs/1805.09573}{{\tt arXiv:1805.09573}}.

\bibitem{Almheiri:2013hfa}
A.~Almheiri, D.~Marolf, J.~Polchinski, D.~Stanford, and J.~Sully, {\it {An
  Apologia for Firewalls}},  {\em JHEP} {\bf 09} (2013) 018,
  [\href{http://arxiv.org/abs/1304.6483}{{\tt arXiv:1304.6483}}].

\bibitem{Almheiri:2012rt}
A.~Almheiri, D.~Marolf, J.~Polchinski, and J.~Sully, {\it {Black Holes:
  Complementarity or Firewalls?}},  {\em JHEP} {\bf 02} (2013) 062,
  [\href{http://arxiv.org/abs/1207.3123}{{\tt arXiv:1207.3123}}].

\bibitem{Bousso:2012as}
R.~Bousso, {\it {Complementarity Is Not Enough}},  {\em Phys. Rev. D} {\bf 87}
  (2013), no.~12 124023, [\href{http://arxiv.org/abs/1207.5192}{{\tt
  arXiv:1207.5192}}].

\bibitem{Abedi:2016hgu}
J.~Abedi, H.~Dykaar, and N.~Afshordi, {\it {Echoes from the Abyss: Tentative
  evidence for Planck-scale structure at black hole horizons}},  {\em Phys.
  Rev. D} {\bf 96} (2017), no.~8 082004,
  [\href{http://arxiv.org/abs/1612.00266}{{\tt arXiv:1612.00266}}].

\bibitem{Polchinski:2016hrw}
J.~Polchinski, {\it {The Black Hole Information Problem}},  in {\em
  {Theoretical Advanced Study Institute in Elementary Particle Physics}: {New
  Frontiers in Fields and Strings}}, 9, 2016.
\newblock \href{http://arxiv.org/abs/1609.04036}{{\tt arXiv:1609.04036}}.

\bibitem{Kaplan:2018dqx}
D.~E. Kaplan and S.~Rajendran, {\it {Firewalls in General Relativity}},  {\em
  Phys. Rev. D} {\bf 99} (2019), no.~4 044033,
  [\href{http://arxiv.org/abs/1812.00536}{{\tt arXiv:1812.00536}}].

\bibitem{McManus:2020lgm}
R.~McManus, E.~Berti, D.~E. Kaplan, and S.~Rajendran, {\it {Quasinormal modes
  and stability of firewalls}},  {\em Phys. Rev. D} {\bf 102} (2020), no.~10
  104031, [\href{http://arxiv.org/abs/2007.15525}{{\tt arXiv:2007.15525}}].

\bibitem{Chen:2014jwq}
P.~Chen, Y.~C. Ong, and D.-h. Yeom, {\it {Black Hole Remnants and the
  Information Loss Paradox}},  {\em Phys. Rept.} {\bf 603} (2015) 1--45,
  [\href{http://arxiv.org/abs/1412.8366}{{\tt arXiv:1412.8366}}].

\bibitem{Braunstein07}
S.~L. Braunstein and A.~K. Pati, {\it {Quantum information cannot be completely
  hidden in correlations: Implications for the black-hole information
  paradox}},  {\em Phys. Rev. Lett.} {\bf 98} (2007) 080502,
  [\href{http://arxiv.org/abs/gr-qc/0603046}{{\tt gr-qc/0603046}}].

\bibitem{Bousso:2020kmy}
R.~Bousso and E.~Wildenhain, {\it {Gravity/ensemble duality}},  {\em Phys. Rev.
  D} {\bf 102} (2020), no.~6 066005,
  [\href{http://arxiv.org/abs/2006.16289}{{\tt arXiv:2006.16289}}].

\bibitem{Giddings:2006vu}
S.~B. Giddings, {\it {Locality in quantum gravity and string theory}},  {\em
  Phys. Rev. D} {\bf 74} (2006) 106006,
  [\href{http://arxiv.org/abs/hep-th/0604072}{{\tt hep-th/0604072}}].

\bibitem{wigner1995remarks}
E.~P. Wigner, {\it Remarks on the mind-body question},  in {\em Philosophical
  reflections and syntheses}, pp.~247--260.
\newblock Springer, 1995.

\bibitem{deutsch1985quantum}
D.~Deutsch, {\it Quantum theory as a universal physical theory},  {\em
  International Journal of Theoretical Physics} {\bf 24} (1985), no.~1 1--41.

\bibitem{bong2020strong}
K.-W. Bong, A.~Utreras-Alarc{\'o}n, F.~Ghafari, Y.-C. Liang, N.~Tischler, E.~G.
  Cavalcanti, G.~J. Pryde, and H.~M. Wiseman, {\it A strong no-go theorem on
  the wigner’s friend paradox},  {\em Nature Physics} {\bf 16} (2020), no.~12
  1199--1205.

\bibitem{schlosshauer2005decoherence}
M.~Schlosshauer, {\it Decoherence, the measurement problem, and interpretations
  of quantum mechanics},  {\em Reviews of Modern physics} {\bf 76} (2005),
  no.~4 1267.

\bibitem{Bousso:2013uka}
R.~Bousso and D.~Stanford, {\it {Measurements without Probabilities in the
  Final State Proposal}},  {\em Phys. Rev. D} {\bf 89} (2014), no.~4 044038,
  [\href{http://arxiv.org/abs/1310.7457}{{\tt arXiv:1310.7457}}].

\bibitem{Pasterski:2020xvn}
S.~Pasterski and H.~Verlinde, {\it {HPS meets AMPS: How Soft Hair Dissolves the
  Firewall}},  \href{http://arxiv.org/abs/2012.03850}{{\tt arXiv:2012.03850}}.

\bibitem{Giddings:2004ud}
S.~B. Giddings and M.~Lippert, {\it {The Information paradox and the locality
  bound}},  {\em Phys. Rev. D} {\bf 69} (2004) 124019,
  [\href{http://arxiv.org/abs/hep-th/0402073}{{\tt hep-th/0402073}}].

\bibitem{Giddings:2005id}
S.~B. Giddings, D.~Marolf, and J.~B. Hartle, {\it {Observables in effective
  gravity}},  {\em Phys. Rev. D} {\bf 74} (2006) 064018,
  [\href{http://arxiv.org/abs/hep-th/0512200}{{\tt hep-th/0512200}}].

\bibitem{kofler2007classical}
J.~Kofler and {\v{C}}.~Brukner, {\it Classical world arising out of quantum
  physics under the restriction of coarse-grained measurements},  {\em Physical
  Review Letters} {\bf 99} (2007), no.~18 180403,
  [\href{http://arxiv.org/abs/quant-ph/0609079}{{\tt quant-ph/0609079}}].

\bibitem{penrose1996gravity}
R.~Penrose, {\it On gravity's role in quantum state reduction},  {\em General
  relativity and gravitation} {\bf 28} (1996), no.~5 581--600.

\bibitem{bassi2013models}
A.~Bassi, K.~Lochan, S.~Satin, T.~P. Singh, and H.~Ulbricht, {\it Models of
  wave-function collapse, underlying theories, and experimental tests},  {\em
  Reviews of Modern Physics} {\bf 85} (2013), no.~2 471.

\bibitem{Belinsky:1970ew}
V.~A. Belinsky, I.~M. Khalatnikov, and E.~M. Lifshitz, {\it {Oscillatory
  approach to a singular point in the relativistic cosmology}},  {\em Adv.
  Phys.} {\bf 19} (1970) 525--573.

\bibitem{Harlow:2014yka}
D.~Harlow, {\it {Jerusalem Lectures on Black Holes and Quantum Information}},
  {\em Rev. Mod. Phys.} {\bf 88} (2016) 015002,
  [\href{http://arxiv.org/abs/1409.1231}{{\tt arXiv:1409.1231}}].

\bibitem{Papadodimas:2013kwa}
K.~Papadodimas and S.~Raju, {\it {The unreasonable effectiveness of
  exponentially suppressed corrections in preserving information}},  {\em Int.
  J. Mod. Phys. D} {\bf 22} (2013) 1342030.

\bibitem{Papadodimas:2012aq}
K.~Papadodimas and S.~Raju, {\it {An Infalling Observer in AdS/CFT}},  {\em
  JHEP} {\bf 10} (2013) 212, [\href{http://arxiv.org/abs/1211.6767}{{\tt
  arXiv:1211.6767}}].

\bibitem{Stanford:2020wkf}
D.~Stanford, {\it {More quantum noise from wormholes}},
  \href{http://arxiv.org/abs/2008.08570}{{\tt arXiv:2008.08570}}.

\bibitem{NC10}
M.~A. Nielsen and I.~L. Chuang, {\it Quantum computation and quantum
  information},  2010.

\bibitem{Bennett93}
C.~H. Bennett, G.~Brassard, C.~Cr{\'e}peau, R.~Jozsa, A.~Peres, and W.~K.
  Wootters, {\it Teleporting an unknown quantum state via dual classical and
  einstein-podolsky-rosen channels},  {\em Physical Review Letters} {\bf 70}
  (1993), no.~13 1895.

\bibitem{Wootters82}
W.~K. Wootters and W.~H. Zurek, {\it A single quantum cannot be cloned},  {\em
  Nature} {\bf 299} (1982), no.~5886 802--803.

\bibitem{schrodinger35}
E.~Schr{\"o}dinger, {\it Discussion of probability relations between separated
  systems},  in {\em Mathematical Proceedings of the Cambridge Philosophical
  Society}, vol.~31, pp.~555--563, Cambridge University Press, 1935.

\bibitem{Uola20}
R.~Uola, A.~C. Costa, H.~C. Nguyen, and O.~G{\"u}hne, {\it Quantum steering},
  {\em Reviews of Modern Physics} {\bf 92} (2020), no.~1 015001,
  [\href{http://arxiv.org/abs/1903.06663}{{\tt arXiv:1903.06663}}].

\bibitem{Bouwmeester97}
D.~Bouwmeester, J.-W. Pan, K.~Mattle, M.~Eibl, H.~Weinfurter, and A.~Zeilinger,
  {\it Experimental quantum teleportation},  {\em Nature} {\bf 390} (1997),
  no.~6660 575--579.

\bibitem{Pan98}
J.-W. Pan, D.~Bouwmeester, H.~Weinfurter, and A.~Zeilinger, {\it Experimental
  entanglement swapping: entangling photons that never interacted},  {\em
  Physical Review Letters} {\bf 80} (1998), no.~18 3891.

\bibitem{Herbst15}
T.~Herbst, T.~Scheidl, M.~Fink, J.~Handsteiner, B.~Wittmann, R.~Ursin, and
  A.~Zeilinger, {\it Teleportation of entanglement over 143 km},  {\em
  Proceedings of the National Academy of Sciences} {\bf 112} (2015), no.~46
  14202--14205, [\href{http://arxiv.org/abs/1403.0009}{{\tt arXiv:1403.0009}}].

\bibitem{ma2012experimental}
X.-s. Ma, S.~Zotter, J.~Kofler, R.~Ursin, T.~Jennewein, {\v{C}}.~Brukner, and
  A.~Zeilinger, {\it Experimental delayed-choice entanglement swapping},  {\em
  Nature Physics} {\bf 8} (2012), no.~6 479--484.

\bibitem{Ma:2014dta}
X.-s. Ma, J.~Kofler, and A.~Zeilinger, {\it {Delayed-choice gedanken
  experiments and their realizations}},  {\em Rev. Mod. Phys.} {\bf 88} (2016)
  015005, [\href{http://arxiv.org/abs/1407.2930}{{\tt arXiv:1407.2930}}].

\bibitem{Hayden:2007cs}
P.~Hayden and J.~Preskill, {\it {Black holes as mirrors: Quantum information in
  random subsystems}},  {\em JHEP} {\bf 09} (2007) 120,
  [\href{http://arxiv.org/abs/0708.4025}{{\tt arXiv:0708.4025}}].

\bibitem{Sekino:2008he}
Y.~Sekino and L.~Susskind, {\it {Fast Scramblers}},  {\em JHEP} {\bf 10} (2008)
  065, [\href{http://arxiv.org/abs/0808.2096}{{\tt arXiv:0808.2096}}].

\bibitem{Hamilton:2006az}
A.~Hamilton, D.~N. Kabat, G.~Lifschytz, and D.~A. Lowe, {\it {Holographic
  representation of local bulk operators}},  {\em Phys. Rev. D} {\bf 74} (2006)
  066009, [\href{http://arxiv.org/abs/hep-th/0606141}{{\tt hep-th/0606141}}].

\bibitem{Czech:2012bh}
B.~Czech, J.~L. Karczmarek, F.~Nogueira, and M.~Van~Raamsdonk, {\it {The
  Gravity Dual of a Density Matrix}},  {\em Class. Quant. Grav.} {\bf 29}
  (2012) 155009, [\href{http://arxiv.org/abs/1204.1330}{{\tt
  arXiv:1204.1330}}].

\bibitem{Kallosh:1995hi}
R.~Kallosh, A.~D. Linde, D.~A. Linde, and L.~Susskind, {\it {Gravity and global
  symmetries}},  {\em Phys. Rev. D} {\bf 52} (1995) 912--935,
  [\href{http://arxiv.org/abs/hep-th/9502069}{{\tt hep-th/9502069}}].

\bibitem{Hsin:2020mfa}
P.-S. Hsin, L.~V. Iliesiu, and Z.~Yang, {\it {A violation of global symmetries
  from replica wormholes and the fate of black hole remnants}},
  \href{http://arxiv.org/abs/2011.09444}{{\tt arXiv:2011.09444}}.

\bibitem{Dong:2016eik}
X.~Dong, D.~Harlow, and A.~C. Wall, {\it {Reconstruction of Bulk Operators
  within the Entanglement Wedge in Gauge-Gravity Duality}},  {\em Phys. Rev.
  Lett.} {\bf 117} (2016), no.~2 021601,
  [\href{http://arxiv.org/abs/1601.05416}{{\tt arXiv:1601.05416}}].

\bibitem{Maldacena:2013xja}
J.~Maldacena and L.~Susskind, {\it Cool horizons for entangled black holes},
  {\em Fortschritte der Physik} {\bf 61} (2013), no.~9 781--811,
  [\href{http://arxiv.org/abs/1306.0533}{{\tt arXiv:1306.0533}}].

\bibitem{Penington:2019kki}
G.~Penington, S.~H. Shenker, D.~Stanford, and Z.~Yang, {\it {Replica wormholes
  and the black hole interior}},  \href{http://arxiv.org/abs/1911.11977}{{\tt
  arXiv:1911.11977}}.

\bibitem{Pollack:2020gfa}
J.~Pollack, M.~Rozali, J.~Sully, and D.~Wakeham, {\it {Eigenstate
  Thermalization and Disorder Averaging in Gravity}},  {\em Phys. Rev. Lett.}
  {\bf 125} (2020), no.~2 021601, [\href{http://arxiv.org/abs/2002.02971}{{\tt
  arXiv:2002.02971}}].

\bibitem{Lee:2020aft}
D.~J. Lee and D.-h. Yeom, {\it {Almost certain loss from black holes: critical
  comments on the black hole final state proposal}},
  \href{http://arxiv.org/abs/2009.08565}{{\tt arXiv:2009.08565}}.

\bibitem{Page:1993wv}
D.~N. Page, {\it {Information in black hole radiation}},  {\em Phys. Rev.
  Lett.} {\bf 71} (1993) 3743--3746,
  [\href{http://arxiv.org/abs/hep-th/9306083}{{\tt hep-th/9306083}}].

\bibitem{Page:1993df}
D.~N. Page, {\it {Average entropy of a subsystem}},  {\em Phys. Rev. Lett.}
  {\bf 71} (1993) 1291--1294, [\href{http://arxiv.org/abs/gr-qc/9305007}{{\tt
  gr-qc/9305007}}].

\bibitem{Perry:2021mch}
M.~J. Perry, {\it {No Future in Black Holes}},
  \href{http://arxiv.org/abs/2106.03715}{{\tt arXiv:2106.03715}}.

\bibitem{Reeh:1961ujh}
H.~Reeh and S.~Schlieder, {\it {Bemerkungen zur unit\"ar\"aquivalenz von
  lorentzinvarianten feldern}},  {\em Nuovo Cim.} {\bf 22} (1961), no.~5
  1051--1068.

\bibitem{Witten:2018zxz}
E.~Witten, {\it {APS Medal for Exceptional Achievement in Research: Invited
  article on entanglement properties of quantum field theory}},  {\em Rev. Mod.
  Phys.} {\bf 90} (2018), no.~4 045003,
  [\href{http://arxiv.org/abs/1803.04993}{{\tt arXiv:1803.04993}}].

\bibitem{tHooft:1993dmi}
G.~'t~Hooft, {\it {Dimensional reduction in quantum gravity}},  {\em Conf.
  Proc. C} {\bf 930308} (1993) 284--296,
  [\href{http://arxiv.org/abs/gr-qc/9310026}{{\tt gr-qc/9310026}}].

\bibitem{Susskind:1994vu}
L.~Susskind, {\it {The World as a hologram}},  {\em J. Math. Phys.} {\bf 36}
  (1995) 6377--6396, [\href{http://arxiv.org/abs/hep-th/9409089}{{\tt
  hep-th/9409089}}].

\bibitem{Wang:2021mqq}
X.~Wang, R.~Li, and J.~Wang, {\it {Page curves for a family of exactly solvable
  evaporating black holes}},  {\em Phys. Rev. D} {\bf 103} (2021), no.~12
  126026, [\href{http://arxiv.org/abs/2104.00224}{{\tt arXiv:2104.00224}}].

\bibitem{Hartman:2020swn}
T.~Hartman, E.~Shaghoulian, and A.~Strominger, {\it {Islands in Asymptotically
  Flat 2D Gravity}},  {\em JHEP} {\bf 07} (2020) 022,
  [\href{http://arxiv.org/abs/2004.13857}{{\tt arXiv:2004.13857}}].

\bibitem{Gautason:2020tmk}
F.~F. Gautason, L.~Schneiderbauer, W.~Sybesma, and L.~Thorlacius, {\it {Page
  Curve for an Evaporating Black Hole}},  {\em JHEP} {\bf 05} (2020) 091,
  [\href{http://arxiv.org/abs/2004.00598}{{\tt arXiv:2004.00598}}].

\bibitem{Gottesman99}
D.~Gottesman and I.~L. Chuang, {\it Demonstrating the viability of universal
  quantum computation using teleportation and single-qubit operations},  {\em
  Nature} {\bf 402} (1999), no.~6760 390--393.

\bibitem{Zhang13}
Y.~Zhang and J.~Pang, {\it Space-time topology in teleportation-based quantum
  computation},  {\em arXiv preprint arXiv:1309.0955} (2013).

\bibitem{Zhang16}
Y.~Zhang, K.~Zhang, and J.~Pang, {\it Teleportation-based quantum computation,
  extended temperley--lieb diagrammatical approach and yang--baxter equation},
  {\em Quantum Information Processing} {\bf 15} (2016), no.~1 405--464.

\bibitem{Temperley71}
H.~N. Temperley and E.~H. Lieb, {\it Relations between the ‘percolation’and
  ‘colouring’problem and other graph-theoretical problems associated with
  regular planar lattices: some exact results for the
  ‘percolation’problem},  {\em Proceedings of the Royal Society of London.
  A. Mathematical and Physical Sciences} {\bf 322} (1971), no.~1549 251--280.

\bibitem{Diosi:1986nu}
L.~Diosi, {\it {A Universal Master Equation for the Gravitational Violation of
  Quantum Mechanics}},  {\em Phys. Lett. A} {\bf 120} (1987) 377.

\bibitem{Diosi:1988uy}
L.~Diosi, {\it {MODELS FOR UNIVERSAL REDUCTION OF MACROSCOPIC QUANTUM
  FLUCTUATIONS}},  {\em Phys. Rev. A} {\bf 40} (1989) 1165--1174.

\bibitem{Donadi:2020kzc}
S.~Donadi, K.~Piscicchia, C.~Curceanu, L.~Di\'osi, M.~Laubenstein, and
  A.~Bassi, {\it {Underground test of gravity-related wave function collapse}},
   {\em Nature Phys.} {\bf 17} (2021), no.~1 74--78.

\bibitem{Wang:2021woy}
X.~Wang, R.~Li, and J.~Wang, {\it {Islands and Page curves of
  Reissner-Nordstr\"om black holes}},  {\em JHEP} {\bf 04} (2021) 103,
  [\href{http://arxiv.org/abs/2101.06867}{{\tt arXiv:2101.06867}}].

\bibitem{Hashimoto:2020cas}
K.~Hashimoto, N.~Iizuka, and Y.~Matsuo, {\it {Islands in Schwarzschild black
  holes}},  {\em JHEP} {\bf 06} (2020) 085,
  [\href{http://arxiv.org/abs/2004.05863}{{\tt arXiv:2004.05863}}].

\bibitem{Li:2021lfo}
R.~Li, X.~Wang, and J.~Wang, {\it {Island may not save the information paradox
  of Liouville black holes}},  \href{http://arxiv.org/abs/2105.03271}{{\tt
  arXiv:2105.03271}}.

\bibitem{Sybesma:2020fxg}
W.~Sybesma, {\it {Pure de Sitter space and the island moving back in time}},
  {\em Class. Quant. Grav.} {\bf 38} (2021), no.~14 145012,
  [\href{http://arxiv.org/abs/2008.07994}{{\tt arXiv:2008.07994}}].

\bibitem{tHooft:2018zwd}
G.~'t~Hooft, {\it {What happens in a black hole when a particle meets its
  antipode}},  \href{http://arxiv.org/abs/1804.05744}{{\tt arXiv:1804.05744}}.

\bibitem{Hawking:2016msc}
S.~W. Hawking, M.~J. Perry, and A.~Strominger, {\it {Soft Hair on Black
  Holes}},  {\em Phys. Rev. Lett.} {\bf 116} (2016), no.~23 231301,
  [\href{http://arxiv.org/abs/1601.00921}{{\tt arXiv:1601.00921}}].

\end{thebibliography}
\end{document}